\documentclass{aa}
\usepackage{latexsym}
\usepackage{rotating}
\usepackage{graphics}
\usepackage{times}
\newcommand{\srm}{\scriptscriptstyle \rm}

\newcommand{\vmax}{v_{\rm max}}

\newcommand{\civ}  {\ion{C}{iv}}
\newcommand{\ciii} {\ion{C}{iii]}}
\newcommand{\oiii} {\ion{[O}{iii]}}
\newcommand{\nv}   {\ion{N}{v}}
\newcommand{\siiv} {\ion{Si}{iv}}
\newcommand{\mgii} {\ion{Mg}{ii}}
\newcommand{\feii} {\ion{Fe}{ii}}
\newcommand{\feiii} {\ion{Fe}{iii}}
\newcommand{\aliii}{\ion{Al}{iii}}
\begin{document}
\title{Polarization properties of broad absorption line QSOs : 
       new statistical clues 
\thanks{Based on observations collected at the European Southern Observatory 
(ESO, La Silla)},
\thanks{Table 1 is also available in electronic form at the CDS via 
http://cdsweb.u-strasbg.fr}}
\author{H. Lamy\inst{1} \and D. Hutsem\'ekers\inst{2,}\thanks{Also, Chercheur 
  Qualifi\'e au Fonds National de la Recherche Scientifique (FNRS, Belgium)}}
\institute{IASB-BIRA, Avenue Circulaire 3, 1180 Bruxelles, Belgium \and 
Institut d'Astrophysique, Universit\'e de Li\`ege, Sart-Tilman, B-4000 
Li\`ege, Belgium}
\offprints{Lamy, \email{herve.lamy@oma.be}}
\date{Received ; accepted }
\abstract{We report the results of several statistical tests performed
on a large sample of 139 broad absorption line (BAL) QSOs with
good quality optical spectra and/or optical polarization data.
Correlations between ten optical indices and the polarization degree
$p_0$ are systematically searched for. We find six significant
non-trivial correlations.  In order to identify the most important
correlations, we perform a principal component analysis with a sample
of 30 BAL QSOs and eight quantities (including $p_0$).  Most of the
variance ($\sim 57\,\%$) in the data is contained in two
principal components called PC1 and PC2.  PC1 is mainly dominated by
the correlation between the balnicity index BI and the strength of the
\feii\, emission; it may be driven by the accretion rate of matter
onto the central compact object.  The variance in PC2 is essentially
due to the anti-correlation between $p_0$ and the detachment index DI,
indicating that BAL QSOs with P Cygni profiles (DI $\ll$) are usually
more polarized than those objects with \civ\, absorption troughs well
detached from the corresponding emission lines (DI $\gg$).  We show
that PC2 may be related to the orientation of the BAL QSOs with
respect to the line of sight.

We also present new spectropolarimetric observations of six BAL QSOs.
By adding spectropolarimetric data from the literature, we build a
sample of 21 BAL QSOs for which we define four spectropolarimetric
indices describing the polarization properties of the absorption and
emission lines.  We find that the polarization of the \ciii\,
emission line is systematically higher than the polarization of the
\civ\, emission line, and that the highest polarization in the troughs is
correlated to the balnicity index.  Another important result emerging
from the statistical tests performed on this spectropolarimetric
sample is a possible anti-correlation between the detachment index and
a quantity SI which measures the ratio of the depths of the \civ\,
absorption in the polarized flux and in the total flux.  This
correlation indicates that in BAL QSOs with P cygni profiles, the
BAL troughs in the polarized flux are nearly as deep as in the total
flux while, in BAL QSOs with detached absorptions, the BAL
troughs in the polarized flux are much weaker than in the total
flux.

We show that our main results may be explained in the framework of a
``two-component'' wind model which is a natural extension of the
classical wind-from-disk models.  In this model, the broad absorption
occurs in a dense equatorial wind emerging from the accretion disk,
while scattering and polarization mainly take place in a polar region.
The orientation relative to the observer drives the correlations $p_0$ --
DI and DI -- SI. While most of our observations can be explained within
this framework, there are also several indications that other
polarization mechanisms, and more particularly resonance scattering,
may also be at work.

}
\titlerunning{Polarization properties of broad absorption line QSOs : new
statistical clues}
\authorrunning{Lamy \& Hutsem\'ekers}
\maketitle

\section{Introduction}

About 10 to 20\,\% of optically-selected QSOs exhibit broad absorption
lines (BAL) in the resonance lines of highly ionized species such as
\nv\,$\lambda\,1240$, \civ\,$\lambda\,1549$ or \siiv\,$\lambda\,1397$ 
(Hewett \& Foltz \cite{HEW03}, Tolea et al. \cite{TOL02}, Reichard et al. 
\cite{REI03}).  A smaller fraction ($\sim$ 1\,\%) has in addition broad 
absorption lines from lower ionization species like \mgii\,$\lambda\,2798$ or 
\aliii\,$\lambda\,1858$.  These objects are often referred to as LoBAL or LIBAL
QSOs in the literature, in contrast with those objects with only high
ionization BALs, hereafter denoted HIBAL QSOs\footnote{In order to be
consistent with the notations introduced in our previous papers, we use here the 
acronyms HIBAL-LIBAL QSO instead of the more widely used HiBAL-LoBAL QSO
dichotomy.  The differences are explained in Sect. 2.2}. The broad absorption 
lines are blueshifted relative to the emission lines, indicating the presence 
of outflows with velocities ranging up to $\sim 0.2\,c$.

Despite of more than three decades of intense observation and
modelling, the fundamental nature of the BAL QSOs remains
controversial. Indeed, the physical mechanism that drives the flow 
is still a matter of debate although there are some indications that 
the outflows are driven by radiation pressure on spectral lines and
originate from an accretion disk rotating around a massive black hole 
(Murray et al. \cite{MUR95}, Arav \cite{ARA97}, Proga et al. \cite{PRO00}).  
Acceleration of the outflow by other mechanisms such as gas pressure (Begelman et al. 
\cite{BEG91}), magnetic fields (de Kool \& Begelman \cite{DEK95}) or radiation pressure 
by dust (Voit et al. \cite{VOI93}, Scoville \& Norman \cite{SCO95}, Everett et al. 
\cite{EVE02}) has also been proposed and may contribute as well.  Moreover, it is not 
yet understood whether BAL QSOs and non-BAL QSOs are completely different types of quasars 
or if their properties mainly depend on their orientation towards us.  

In order to determine if BAL QSOs and non-BAL QSOs are drawn from the
same parent population of quasars, Weymann et al. (\cite{WEY91}, hereafter
W91) carried out a detailed comparative study of the emission lines and
continuum properties of a sample of 42 BAL QSOs and 29 non-BAL QSOs.  Recently, 
Reichard et al. (\cite{REI03}) made a similar
analysis with a much larger sample (224 BAL QSOs and 892 non-BAL QSOs) drawn
from the SDSS EDR (Sloan Digital Sky Survey Early Data Release, Stoughton 
et al. \cite{STO02}), which essentially confirms the results of W91 : there are no
significant differences between emission line properties of BAL QSOs and non-BAL QSOs, 
except a noticeable enhancement of \nv \, emission in BAL QSOs.  Also, the continua
of the BAL QSOs are redder (Reichard et al. \cite{REI03}), especially for those objects with
low ionization absorption lines.  This is possibly due to a larger dust content 
(Sprayberry \& Foltz \cite{SPR92}, Yamamoto \& Vansevicius \cite{YAM99}).  Both
statistical studies lead to conclusion that BAL QSOs and non-BAL QSOs
come from a single population.  The small differences observed in their
emission line/continuum properties may indicate that BAL QSOs are normal 
QSOs with extreme properties such as larger accretion rates 
(Reichard et al. \cite{REI03}, Yuan \& Wills \cite{YUA03}).

Partly based on these results, the most popular paradigm considers
that a broad absorption line region (BALR) exists in all QSOs and has
a covering factor $\sim 0.1$, i.e. close to the rate of detection of BAL 
QSOs among optically-selected QSOs.  In this orientation scheme, broad
absorption lines are observed only when our line of sight intercepts
the outflow.  A completely different possibility is to consider that
the BAL phenomenon is an evolutionary phase with significant mass loss 
during the life of a QSO (Voit, Weymann \& Korista \cite{VOI93}).  The best 
argument supporting this evolutionary model is given by the large percentage of 
LIBAL QSOs detected among infrared surveys (Low et al. \cite{LOW89}), suggesting 
that these objects are enshrouded in a cocoon of dust which heavily 
attenuates the optical/UV continuum (Sprayberry \& Foltz \cite{SPR92}).  Moreover, 
in order to explain the very weak \oiii\, intensity measured in those LIBAL 
QSOs (Boroson \& Green \cite{BOR92}, Yuan \& Wills \cite{YUA03}), the covering 
factor of the BALR must be much larger than 0.1.

Since we want to test for orientation as a possible key-parameter to
understand the BAL QSO/non-BAL QSO dichotomy, polarization may be a useful 
tool because it is sensitive to the geometry and orientation of these unresolved 
objects.  Previous optical polarization surveys have shown that, as a class, BAL 
QSOs are more polarized than non-BAL QSOs and that the polarization is most
probably due to scattering (Moore \& Stockman \cite{MOO84}, Hutsem\'ekers et al. \cite{HUT98} 
(hereafter Paper I), Schmidt \& Hines \cite{SCH99}, Hutsem\'ekers \& Lamy \cite{HUT02}, 
Lamy \cite{LAM03}).  The fact that BAL QSOs are polarized reveals that the scattering 
regions are not spherically symmetric.  Many current theoretical models of BAL QSOs assume 
an axisymmetric geometry, with either an accretion disk (e.g. Murray et al. \cite{MUR95}) or 
an opaque dusty torus (e.g. Schmidt \& Hines \cite{SCH97}) located in the equatorial plane. In 
this equatorial geometry, the BALR emerges either from the accretion disk in 
the form of a continuous flow or it is ablated from the top of the dusty molecular 
torus in the form of dense cloudlets. It is then accelerated to large velocities.  
However, there exists recent models which locate the BALR within a weak jet in the 
polar regions in order to explain the radio-quiet/radio-loud dichotomy and to 
simultaneously relate the BAL phenomenom to the associated absorptions often observed 
in radio-loud QSOs (Punsly \cite{PUN99}, Kuncic \cite{KUN99}).  In this polar geometry, 
the BALR is embedded in a poorly collimated radio jet and is accelerated by the pressure 
of the surrounding medium which also acts to confine the BAL clouds.

Spectropolarimetry has been used to put constrains on the structure of BAL QSOs 
(Glenn et al. \cite{GLE94}, Cohen et al. \cite{COH95}, Goodrich \& Miller \cite{GOO95}, 
Schmidt \& Hines \cite{SCH99}, Ogle et al. \cite{OGL99}).  It has revealed the presence 
of a polarized component filling in the absorption troughs and attributed to a continuum 
component polarized by scattering off electrons or dust.  The scattered and direct rays 
suffer different amounts of absorption which results in a strong increase of the polarization 
in the BAL troughs.  Spectropolarimetry has also been used to determine the relative size 
of the different emission/absorption/scattering regions in BAL QSOS. However, several 
axisymetric configurations are possible to interpret the spectropolarimetric observations 
(Ogle \cite{OGL97}).   Therefore, we have to find other parameters which, in connection to 
the polarization data, may help to discriminate among the different models of BAL QSOs.

In this perspective, we performed in Paper I a systematic search 
of correlations between optical broad-band polarization and several indices
characterizing the optical spectra of BAL QSOs in order to have a  
clue on the nature of the outflows. We find that the main correlations 
involving $p_0$ could be naturally interpreted within the ``wind-from-disk" 
(hereafter WfD) model of Murray et al. (\cite{MUR95}).  However, the size of the sample was 
small (29 objects) and some important but statistically marginal trends 
noticed in Paper I need to be confirmed.  Schmidt \& Hines (\cite{SCH99}) also
report on a marginal correlation between $p_0$ and the strength of the
\civ\, broad absorption lines, which was not detected in Paper I. Since then, a lot of 
good quality polarimetric data of BAL QSOs have been obtained (Lamy \& Hutsem\'ekers 
\cite{LAM00a}, Sluse et al. \cite{SLU04}, Hutsem\'ekers et al. \cite{HUT04}) and the 
existence of a population of radio loud QSOs with intrinsic, BAL-like outflows has 
been confirmed (Becker et al. \cite{BEC00}, \cite{BEC01}, Menou et al. \cite{MEN01}). 
It is therefore important to perform these statistical tests again using a much larger 
sample of BAL QSOs than in Paper I (139 objects instead of 29), in order to better 
identify which spectral characteristics of BAL QSOs are orientation dependent. 
This is the scope of this paper. The fact that sometimes many properties are 
correlated makes the understanding of the physical mechanisms driving them
difficult to discern. The Principal Component Analysis (PCA) is a statistical
method which has proven to be very efficient at isolating the most important
correlations in active galactic nuclei (AGN) (e.g. Boroson \& Green \cite{BOR92}). 
Here the PCA will be used with a BAL QSO sample.

Spectropolarimetry of BAL QSOs with a good signal to noise has been published for a 
large number of BAL QSOs (Ogle et al. \cite{OGL99}).  In combination with our own 
spectropolarimetric data, we define sp ectropolarimetric indices to perform a statistical 
analysis.  The results are then used in combination with previous analyses 
to better constrain the BALR models.

In Sect. 2, we describe the sets of data.  The statistical
analyses are done in Sect. 3. Sect. 4 is devoted to a principal
component analysis of the data in order to identify the variables
that correlate together and to relate them to more fundamental
underlying parameters that drive the correlations.  In Sect. 5, we
present the new spectropolarimetry of 6 BAL QSOs.  Considering
additional data from the literature, spectropolarimetric indices are
defined and correlations with broad band optical polarization and with
optical indices are searched for.  In the last section, we
discuss and interpret the main results in the framework of existing
BAL QSO models.

\section{The BAL QSO data}

The sample considered here is made up of 139 BAL QSOs. It is essentially 
a compilation of 93 BAL QSOs coming from large polarization surveys. 
In addition, we also consider 46 BAL QSOs with good 
quality optical (rest-frame UV) spectra but no polarization data.
The optical polarimetric observations were conducted by us (Paper I, Lamy \& 
Hutsem\'ekers \cite{LAM00a}, Sluse et al. \cite{SLU04}, Hutsem\'ekers et al. 
\cite{HUT04}) and by Ogle et al. (\cite{OGL99}) and Schmidt \& Hines (\cite{SCH99}). 
The spectra considered for the measurement of optical indices are those from the W91 
and Korista et al. (\cite{KOR93}) samples (mainly drawn from the LBQS), and from the 
recent FIRST (Becker et al. \cite{BEC00}, \cite{BEC01}) and SDSS/FIRST (Menou et al. 
\cite{MEN01}) surveys. Apart from these major surveys, we do not consider individually 
studied BAL QSOs. Indeed such objects are not very numerous and are usually observed or 
detected because of their peculiar properties. For the same reason, we do not consider 
the unusual BAL QSOs uncovered by Hall et al. (\cite{HAL02}).

The characteristics of the BAL QSOs considered in this study are reported
in Table 1 and described in detail below.

\subsection{The broad band polarization data}

The broad band polarization data are compiled from our previous
surveys (Paper I, Lamy \& Hutsem\'ekers \cite{LAM00a}), from the survey of
Schmidt \& Hines (\cite{SCH99}) and from the spectropolarimetric atlas of Ogle
et al. (\cite{OGL99}).  When an object is observed several times, we choose
the value of the polarization degree with the lowest uncertainty,
i.e. the smallest $\sigma_p$. Our surveys have been conducted in the V
filter with a typical uncertainty $\simeq$ 0.2 \%. These
uncertainties are adequate for BAL QSO studies since the polarization
of these objects peaks around $\sim$ 1\,\% (Paper I, Lamy \cite{LAM03}). The
measurements of Schmidt \& Hines (\cite{SCH99}), mostly obtained in white
light, have often larger uncertainties ($\sigma_p \simeq 0.4 \%$).  The 
data from Ogle et al. (\cite{OGL99}) are high-quality broad-band polarizations 
(4000-8600 \AA) measured from their spectropolarimetric data, and obtained 
either with the Keck or the Palomar telescopes.  The typical uncertainties
range from $\sim 0.05 \%$ for the Keck data to $\sim 0.15 \%$ for the Palomar
data.  Within the uncertainties, no polarization variability is observed for 
BAL QSOs with multiple observations.  Finally, a few additional polarization
measurements of BAL QSOs from the LBQS and FIRST surveys (Becker et al. \cite{BEC00},
Menou et al. \cite{MEN01}) have recently been secured using the ESO 3.6m telescope
at La Silla (Sluse et al. \cite{SLU04}, Hutsem\'ekers et al. \cite{HUT04}) and are 
added to the sample.

The observed degree of polarization $p$ and its uncertainty $\sigma_p$
are given in Table 1 together with the debiased degree
of polarization $p_0$ according to the Wardle \& Kronberg (\cite{WAR74}) method.
Since most BAL QSOs in Table 1 have a polarization degree
of the order of 1\,\%, we adopt a quality requirement on $p_0$ for all 
the statistical tests performed throughout this paper and involving this 
quantity.  Data with poor signal-to-noise ratio are excluded by considering 
only those objects with $p_0/\sigma_p > 2$ if $p_0 \geq 0.80$ \% or  
with $\sigma_p < 0.40$ \% if $p_0 < 0.80$ \%.

\subsection{The BAL QSO sub-types}

We use the classification described in detail in Paper I.  After a
careful visual inspection of the optical spectra available in the
literature, each BAL QSO has been given a number between 2 and 6,
according to the detection and intensity of the low ionization
troughs.  HIBAL QSO (type 2) are those objects which show broad
absorption lines from highly ionized species but do not have broad
absorption lines from \mgii\ \,nor \aliii.  If \aliii \, is not
observed in absorption and if the optical spectrum does not cover the
\mgii \, spectral region, the BAL QSO is considered as an unclassified
BAL QSO (type 6).  All the objects with either \mgii \, or \aliii \,
broad absorption lines in their spectra are classified as LIBAL QSOs.
According to the strength of the low ionization absorption lines, we
define three sub-categories of LIBAL QSOs : S (strong, type 3), W
(weak, type 4) and M (marginal, type 5).  Note that the term ``LoBAL''
widely spread in the literature essentially corresponds to our type 3.

\subsection{Indices derived from the optical spectra}

We give in Table 1 the following indices
originally defined by W91 to characterize the absorption/emission
features in BAL QSOs : (1) the balnicity index BI which is a modified
velocity equivalent width of the \civ\ BAL, (2) the half-width at
half-maximum HWHM and the rest equivalent width REW of the 
\ciii \,$\lambda\,1909$ emission line and of the red part of the \civ\, 
emission line, (3) the rest equivalent widths of emission features 
related to \feii \, and located near 2070 and 2400 \AA\, and (4) the 
detachment index DI which measures the onset velocity of the
strongest\footnote{In objects with multiple absorption troughs, W91
define the strongest absorption as the one with the largest equivalent
width.}  \civ\ BAL trough in units of the adjacent emission line
half-width, that is the degree of detachment of the absorption line
relative to the emission one.

Apart from the W91 sample, none of these indices are given in the literature 
except BI.  Therefore, as already done in Paper I and in Hutsem\'ekers \& Lamy 
(\cite{HUT00}, hereafter paper II), we have computed these indices whenever
possible by using good quality spectra digitally scanned from the literature.
We follow the prescriptions given in W91 and refer the reader to this 
paper for additional details.
The new measurements make use of spectra published by W91, Korista et
al. (\cite{KOR93}), Ogle et al. (\cite{OGL99}), Becker et al. (\cite{BEC00}, 
\cite{BEC01}), Menou et al. (\cite{MEN01}), and a few spectra from the LBQS 
(Foltz et al. \cite{FOL89}, Chaffee et al. \cite{CHA91}, Hewett et al. \cite{HEW91}). 
For one object (\object{B2240$-$3702}), the indices have been 
measured from our spectra (cf. Fig. \ref{fig:spectropola1}).  All these new
measurements are reported in Table 1 together with
values previously published in Paper I and II.

In Table 1, following W91, we consider as BAL~QSOs those objects 
with a non-zero BI in either \civ\ or \mgii\ (since usually BI (\civ ) $\geq$ BI 
(\mgii )).  But for the consistency of the statistical analysis, only the BI 
measured from \civ\ are considered.  The BI values 
in Table 1 are mainly found in the literature (W91,
Korista et al. \cite{KOR93}, Brotherton et al. \cite{BRO98}, \cite{BRO02}, 
Ogle et al. \cite{OGL99}, Becker et al. \cite{BEC00}, \cite{BEC01}, Menou et al. 
\cite{MEN01}). For the objects considered at the same time in the W91 and the 
Korista samples, we adopt the average of their values.  We have calculated BI only  
for those objects observed at the Palomar Observatory by Ogle et al. (\cite{OGL99}).
 
In addition to these indices, we have also reported in Table 1
the maximum velocity in the \civ\ BAL trough, $\vmax$, which provides an estimate 
of the terminal velocity of the flow. For the BAL QSOs not included in the samples 
of Paper I and II, the values of $\vmax$ are either given by Becker et al. (\cite{BEC00}) 
and Menou et al. (\cite{MEN01}), or they are evaluated from the Korista et al. (\cite{KOR93}), 
Ogle et al. (\cite{OGL99}) and Becker et al. (\cite{BEC01}) spectra, by measuring, from 
the blue to the red, the wavelength at which the absorption first drops 10\% below
the flux level defined by the local continuum.  At velocities higher
than $\sim$25000 km~s$^{-1}$, \civ\, BALs may be contaminated by the
\siiv\, emission line, such that measurements of $\vmax$ become
inaccurate. In a conservative way, we therefore limit
$\vmax$ to 25000 km s$^{-1}$ from the \civ\, emission centro\"{\i}d, even
if larger values are sometimes given by some authors (Becker et al. \cite{BEC00}, 
Menou et al. \cite{MEN01}).  In these cases, 25000 km s$^{-1}$
constitutes a lower limit to the true $\vmax$. Finally, we have measured 
the slope of the continuum, $\alpha_B$, by fitting the continuum blueward 
of \ciii \, and assuming a power-law continuum $F_{\nu} \, \propto \, 
\nu^{-\alpha_B}$ (see Paper I for details).

The errors associated with the measurements of these indices
are difficult to estimate.  They are essentially related to the
positioning of the continuum but they also depend on the structure of
the lines, the nominal redshift, etc \ldots  Although the uncertainty on the 
continuum is partly taken into account in its definition, errors on BI may still 
be large. We estimate them from values published by different authors 
(for example the samples of W91 and Korista et al. \cite{KOR93}) : typical 
uncertainties on BI are around a few hundred ${\rm km\,s^{-1}}$.
The typical errors on $v_{\rm max}$ have been estimated from two
independent series of measurements and are typically around $ 1000 \,
{\rm km\,s^{-1}}$.  The typical error on $\alpha_B$ is $\sim 0.3$, as
evaluated by varying the possible continua (Paper I).  The error
associated with the detachment index DI is typically around 15\,\% from
a comparison between our values and those from W91, although it could
be higher in some cases, especially for those objects with multiple troughs 
or complex absorptions.  Indeed, depending on the continuum, one might 
select different troughs as being the strongest one.  By doing the same 
kind of comparison with W91, we estimate the errors associated
with \civ\, HWHM and \civ\, HREW to be $\sim 10 \%$ and $\sim 20 \%$ respectively .

Finally, we have included in Table 1 the B absolute
magnitude of the objects, $M_B$, which is calculated by following the
prescriptions of Veron \& Veron (\cite{VER00}) (with $H_0 = 50\,{\rm km\,s}^{-1}\,
{\rm Mpc}^{-1}$, $q_0=0$ and an optical spectral index $\alpha=0.3$) and
by applying the ``BAL K-correction" of Stocke et al. (1992).

\begin{figure*}
\resizebox{\hsize}{!}{\rotatebox{0}{\includegraphics{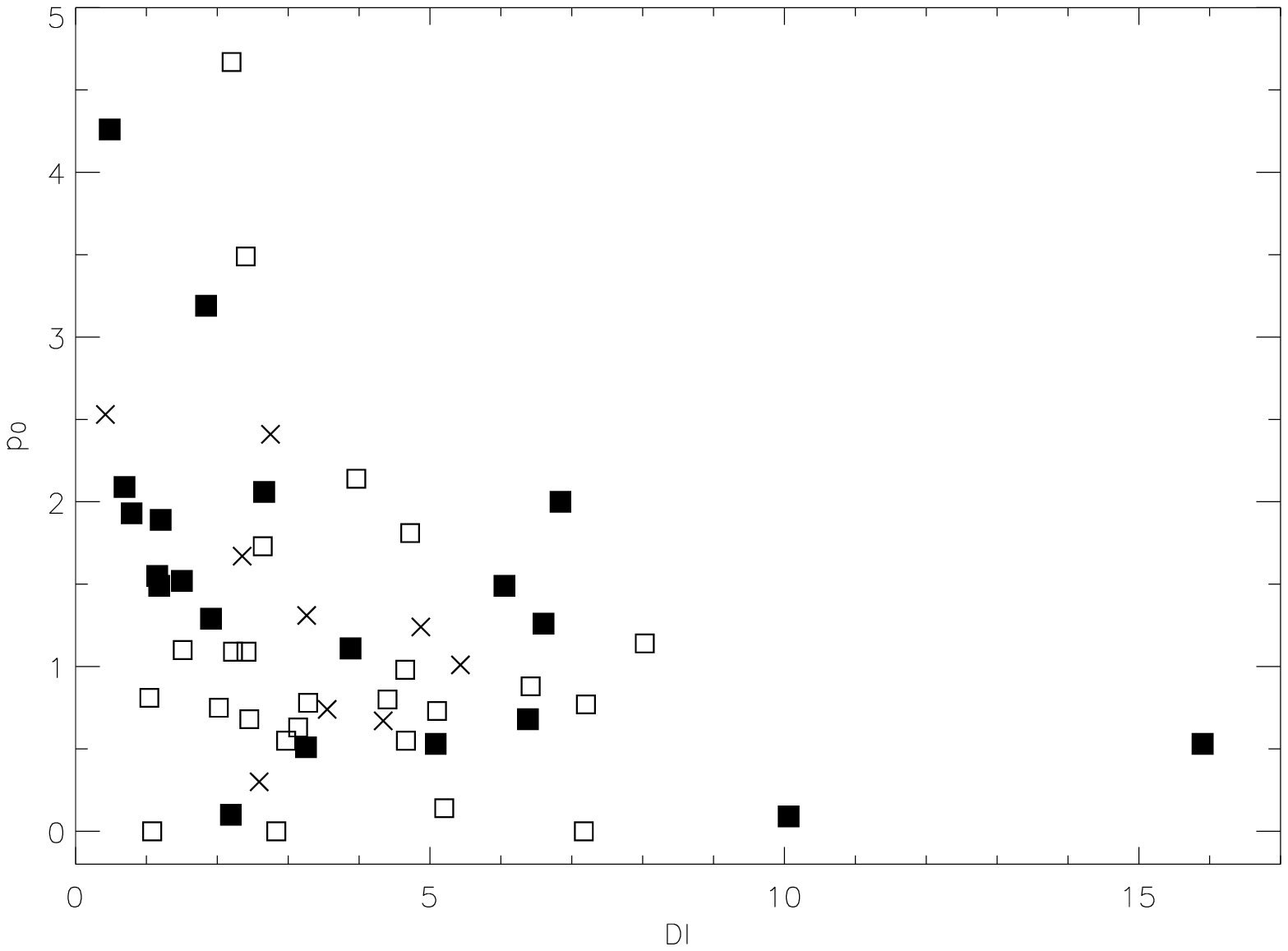}}
{\includegraphics{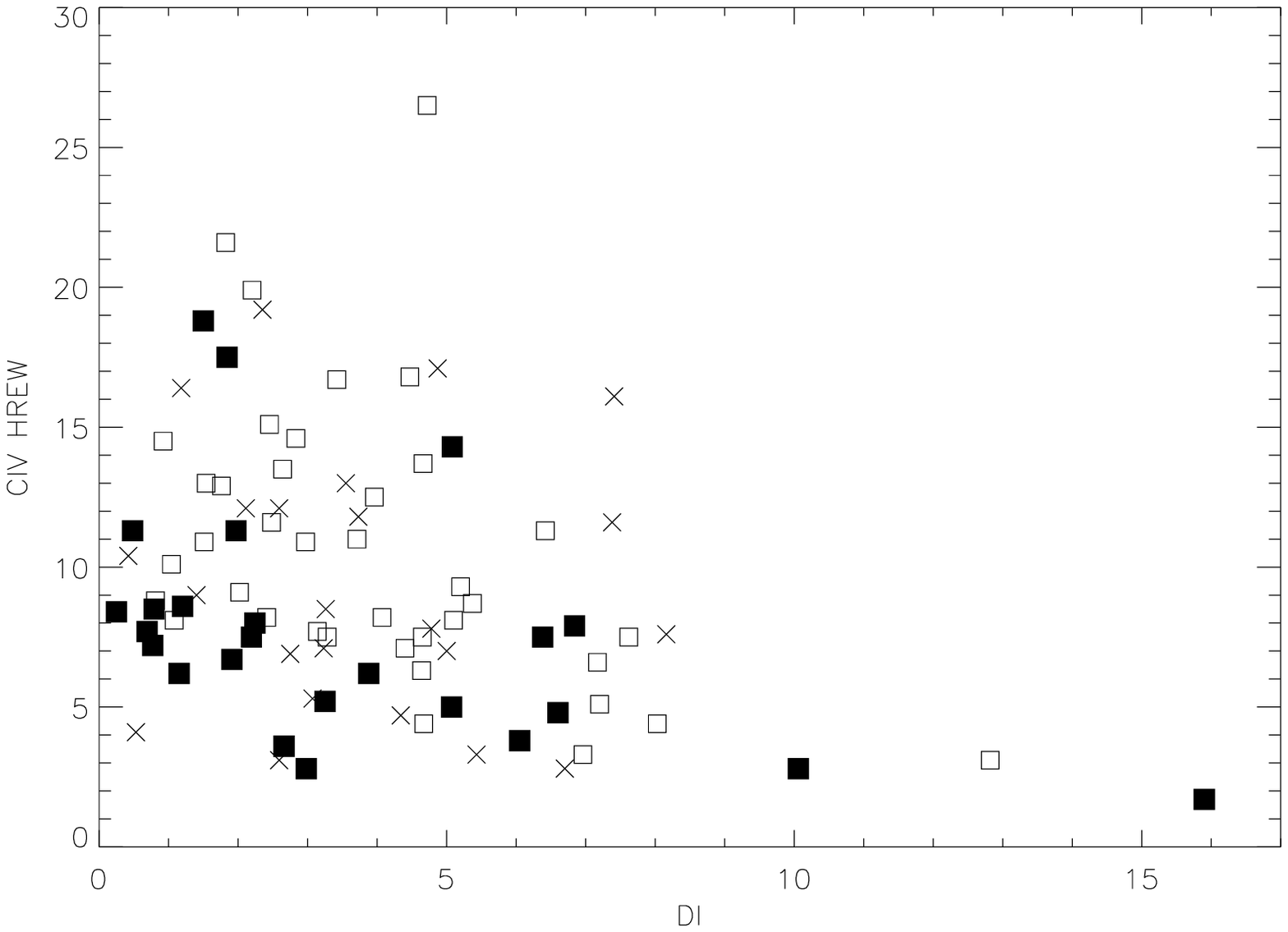}}} \\
\resizebox{\hsize}{!}{\rotatebox{0}{\includegraphics{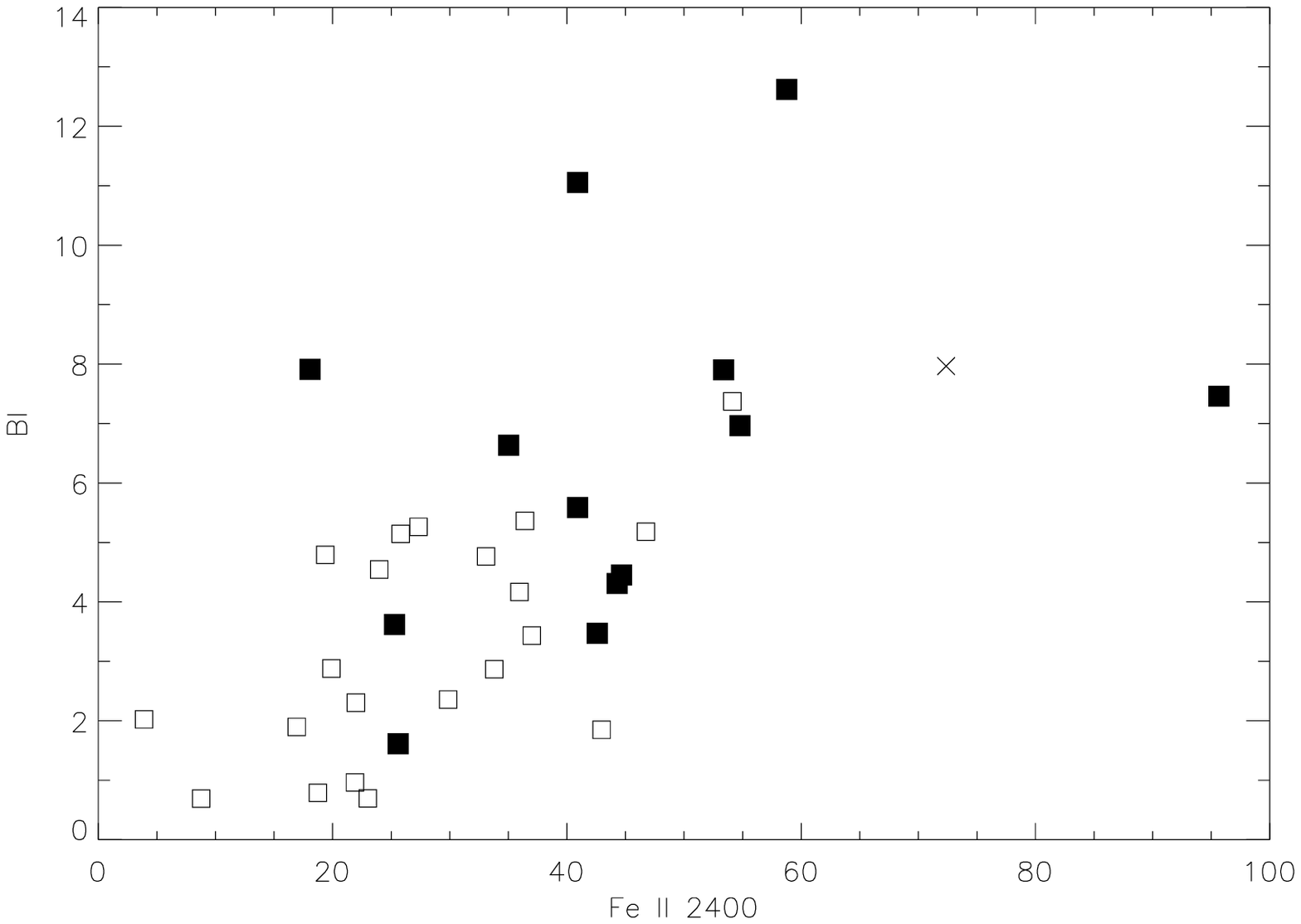}}
{\includegraphics{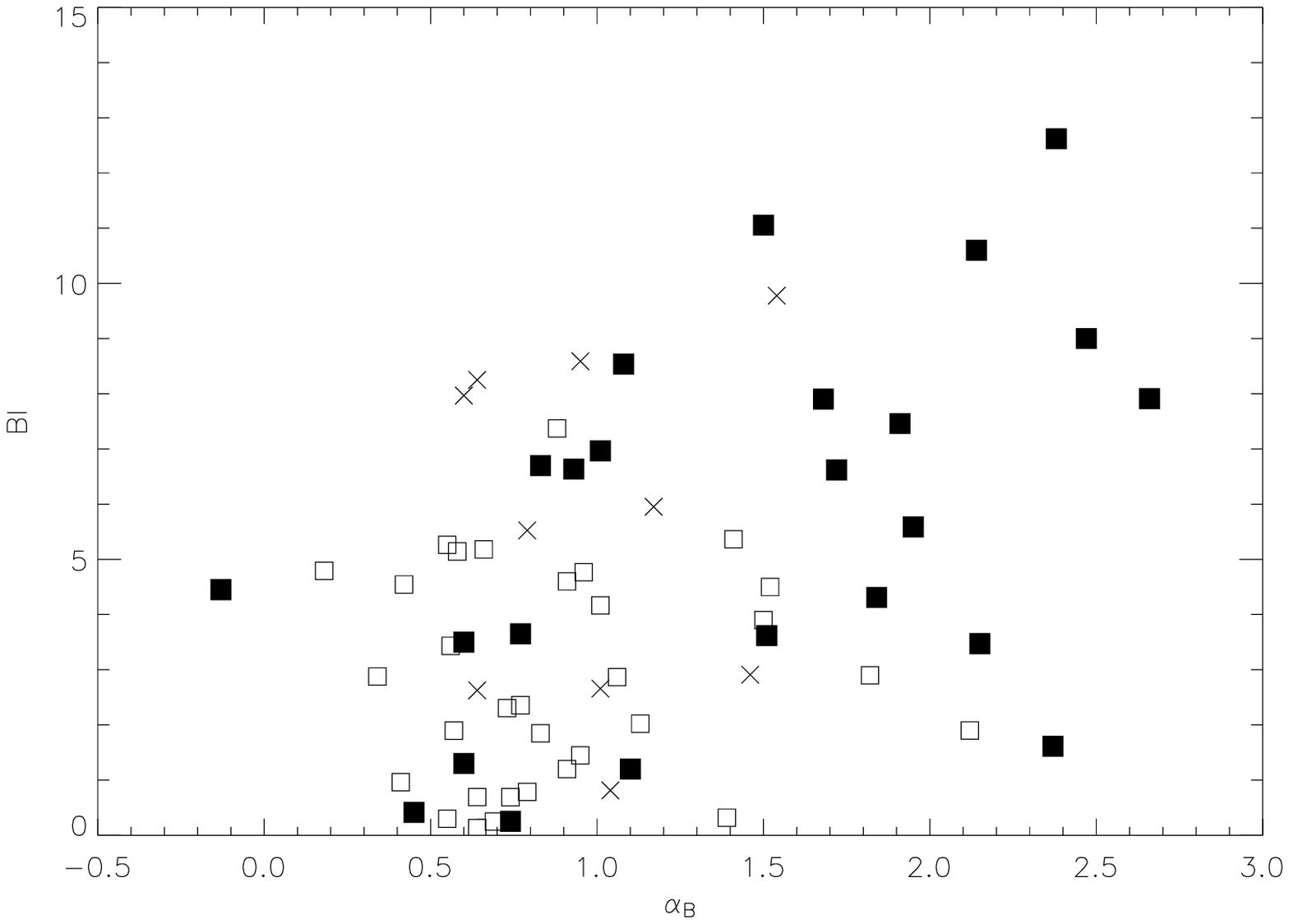}}}
\caption[] {The most significant non-trivial correlations (see text).  
$p_0$ is the debiased broad-band optical polarization (in \%), DI is the detachment index 
(unitless), \civ\, HREW is the rest equivalent width of the red part of the \civ\, BEL (in \AA), 
BI is the balnicity index (in $10^3\,{\rm km\,s}^{-1}$), $\alpha_B$ is the slope of the 
continuum measured shortward of \ciii\, (unitless) and \feii \, 2400 is the rest equivalent
width of the \feii \, emission feature near 2400 \AA. Open squares are HI BAL QSOs, filled squares 
are LI BAL QSOs and crosses are unclassified BAL QSOs.}
\label{fig:corr}
\end{figure*}

\section{Analysis of correlations involving the polarization and spectral 
indices}

Possible correlations involving the optical polarization and the indices 
defined in Sect. 2.3 have been searched for by
computing the Kendall $\tau$ and the Spearman $r_s$ rank correlation
coefficients (Press et al. \cite{PRE89}).  While our main goal is to
investigate the behavior of the polarization, correlations between the
indices themselves have also been searched for since several
new measurements are available.

\setcounter{table}{1}
\begin{table*}[!]
\caption{Analysis of correlations between $p_0$ and the indices described in
Sect. 2.3 as well as correlations between the indices themselves.  Three samples of 
BAL QSOs are considered.  $n$ is the number of objects considered in the tests.
The last column indicates whether the correlation is positive or negative (identical for
the three BAL QSO samples).  Boldfaced entries are the least likely to be due to
random chance.}
\begin{tabular}{llcccccccccc} 
\hline
 &  &  \multicolumn{3}{c}{BAL QSOs} &  \multicolumn{3}{c}{HI BAL QSOs} & \multicolumn{3}{c}{LI BAL QSOs} & \\
Index1 & Index2 & $P_{\tau}$ & $P_{r_s}$ & $n$ & $P_{\tau}$ & $P_{r_s}$ & $n$
& $P_{\tau}$ & $P_{r_s}$ & $n$ & \\
[0.05in] \hline \\[-0.10in]
$p_0$ & BI & 0.176 & 0.180 & 58 & 0.557 & 0.646 & 27 & 0.153 & 0.180 & 21 & $+$\\
$p_0$ & DI & {\bf 0.001} & {\bf 0.003} & 54 & 0.437 & 0.469 & 25 & {\bf 0.004} & {\bf 0.003} & 20 & $-$ \\
$p_0$ & \civ \, ${\srm HWHM}$ & 0.834 & 0.824 & 54 & 0.772 & 0.890 & 26 & 0.888 & 0.898 & 19 & $-$\\
$p_0$ & \civ \, ${\srm HREW}$ & 0.113 & 0.099 & 53 & 0.538 & 0.560 & 25 & 0.078 & 0.046 & 19 & $+$\\
$p_0$ & \ciii \, ${\srm HWHM}$ & 0.059 & 0.085 & 48 & 0.744 & 0.859 & 24 & 0.322 & 0.458 & 16 & $-$ \\
$p_0$ & \ciii \, ${\srm REW}$ & 0.305 &  0.305 & 33 & 0.763 & 0.770 & 15 & 0.170 & 0.308 & 12 & $+$\\
$p_0$ & \feii \, $\lambda\,2400$ & 0.411 & 0.397 & 31 & 0.227 & 0.217 & 17 & 0.663 & 0.792 & 13 & $+$ \\
$p_0$ & \feii \, $\lambda\,2070$ & 0.022 & 0.033 & 38 & {\bf 0.007} & {\bf 0.003} & 17 & 0.516 & 0.541 & 15 & $+$\\
$p_0$ & $\alpha_B$ & 0.082 & 0.081 & 49 & 0.030 & 0.024 & 22 & 0.517 & 0.494 & 18 & $+$\\
$p_0$ & $v_{\rm max}$ & 0.978 & 0.967 & 57 & 0.656 & 0.631 & 27 & 0.710 & 0.627 & 21 & $+$\\
BI    & DI & 0.208 & 0.213 & 93 & 0.022 & 0.036 & 43 & 0.390 & 0.285 & 26 & $+$ \\
BI    & \civ \, ${\srm HWHM}$ & 0.741 & 0.780 & 92 & 0.469 & 0.451 & 43 & 0.513 & 0.583 & 25 & $+$ \\
BI    & \civ \, ${\srm HREW}$ & 0.495 & 0.405 & 90 & 0.965 & 0.869 & 42 & 0.301 & 0.209 & 25 & $-$\\
BI    & \ciii \, ${\srm HWHM}$ & 0.044 & 0.030 & 69 & 0.100 & 0.093 & 41 & 0.384 & 0.409 & 18 & $+$\\
BI    & \ciii \, ${\srm REW}$ & 0.505 & 0.466 & 38 & 0.888 & 0.912 & 19 & 0.273 & 0.354 & 12 & $+$\\
BI    & \feii \, $\lambda\,2070$ & {\bf 0.000} & {\bf 0.000} & 43 & 0.103 & 0.062 & 21 & 0.067 & 0.087 & 15 & $+$\\
BI    & \feii \, $\lambda\,2400$ & {\bf 0.000} & {\bf 0.000} & 35 & {\bf 0.008} & {\bf 0.006} & 21 & 0.197 & 0.254 & 13 & $+$ \\
BI    & $\alpha_B$ & {\bf 0.003} & {\bf 0.003} & 66 & 0.708 & 0.741 & 32 & 0.021 & {\bf 0.016} & 24 & $+$\\
BI    & $v_{\rm max}$ & {\bf 0.000} & {\bf 0.000} & 108 & {\bf 0.000} & {\bf 0.000} & 53 & 0.030 & 0.058 & 30 & $+$ \\ 
DI    & \civ \, ${\srm HWHM}$ & 0.527 & 0.509 & 91 & 0.200 & 0.224 & 42 & 0.513 & 0.619 & 25 & $-$\\
DI    & \civ \, ${\srm HREW}$ & {\bf 0.002} & {\bf 0.001} & 90 & {\bf 0.004} & {\bf 0.001} & 41 & {\bf 0.006} &{\bf 0.004} & 25 & $-$\\
DI    & \ciii \, ${\srm HWHM}$ & 0.247 & 0.217 & 65 & 0.053 & 0.047 & 38 & 0.970 & 0.798 & 18 & $+$\\
DI    & \ciii \, ${\srm REW}$ & 0.040 & 0.054 & 38 & 0.575 & 0.459 & 19 & 0.217 & 0.138 & 12 & $-$\\
DI    & \feii \, $\lambda\,2070$ & 0.335 & 0.310 & 43 & 0.305 & 0.207 & 21 & 0.520 & 0.576 & 15 & $+$ \\
DI    & \feii \, $\lambda\,2400$ & 0.881 & 0.092 & 35 & 0.398 & 0.336 & 21 & 0.243 & 0.343 & 13 & $+$ \\
DI    & $\alpha_B$ & 0.532 & 0.469 & 61 & 0.836 & 0.868 & 29 & 0.843 & 0.922 & 22 & $-$\\
DI    & $v_{\rm max}$ & {\bf 0.000} & {\bf 0.000} & 93 & {\bf 0.000} & {\bf 0.000} & 43 & 0.040 & 0.038 & 26 & $+$\\
\civ \, ${\srm HWHM}$ & \civ \, ${\srm HREW}$ & 0.387 & 0.415 & 90 & 0.616 & 0.678 & 42 & 0.082 & 0.131 & 25 & $+$\\
\civ \, ${\srm HWHM}$ & \ciii \, ${\srm HWHM}$ & {\bf 0.004} & {\bf 0.005} & 64 & 0.105 & 0.148 & 37 & {\bf 0.017} & {\bf 0.011} & 18 & $+$\\
\civ \, ${\srm HWHM}$ & \ciii \, ${\srm REW}$ & 0.546 & 0.589 & 37 & 0.909 & 0.925 & 18 & 1.000 & 0.931 & 12 & $-$\\
\civ \, ${\srm HWHM}$ & \feii \, $\lambda\,2070$ & 0.204 & 0.205 & 41 & 0.192 & 0.229 & 20 & 0.702 & 0.887 & 14 & $+$\\
\civ \, ${\srm HWHM}$ & \feii \, $\lambda\,2400$ & 0.630 & 0.635 & 33 & 0.059 & 0.051 & 20 & 0.131 & 0.199 & 12 & $+$\\
\civ \, ${\srm HWHM}$ & $\alpha_B$ & 0.442 & 0.380 & 59 & 0.858 & 0.843 & 28 & 0.080 & 0.038 & 21 & $+$\\
\civ \, ${\srm HWHM}$ & $v_{\rm max}$ & {\bf 0.014} & {\bf 0.015} & 92 & 0.066 & 0.094 & 43 & 0.228 & 0.221 & 25 & $+$\\
\civ \, ${\srm HREW}$ & \ciii \, ${\srm HWHM}$ & {\bf 0.003} & {\bf 0.011} & 63 & {\bf 0.006} & {\bf 0.012} & 36 & 0.340 & 0.373 & 18 & $-$\\
\civ \, ${\srm HREW}$ & \ciii \, ${\srm REW}$ & {\bf 0.000} & {\bf 0.000} & 37 & 0.027 & 0.029 & 18 & 0.095 & 0.042 & 12 & $+$ \\
\civ \, ${\srm HREW}$ & \feii \, $\lambda\,2070$ & 0.498 & 0.525 & 41 & 0.845 & 0.885 & 20 & 0.868 & 0.976 & 14 & $-$\\
\civ \, ${\srm HREW}$ & \feii \, $\lambda\,2400$ & 0.938 & 0.985 & 33 & 0.896 & 0.830 & 20 & 0.403 & 0.555 & 12 & $-$\\
\civ \, ${\srm HREW}$ & $\alpha_B$ & 0.597 & 0.639 & 58 & 0.983 & 0.882 & 27 & 0.378 & 0.339 & 21 & $-$\\
\civ \, ${\srm HREW}$ & $v_{\rm max}$ & 0.069 & 0.042 & 90 & 0.029 & 0.028 & 42 & 0.905 & 0.830 & 25 & $-$\\
\ciii \, ${\srm HWHM}$ & \ciii \, ${\srm REW}$ & 0.512 & 0.498 & 38 & 0.916 & 0.932 & 19 & 0.272 & 0.183 & 12 & $+$\\
\ciii \, ${\srm HWHM}$ & \feii \, $\lambda\,2070$ & 0.275 & 0.228 & 41 & 0.952 & 0.867 & 21 & 0.088 & 0.061 & 13 & $+$\\
\ciii \, ${\srm HWHM}$ & \feii \, $\lambda\,2400$ & 0.852 & 0.867 & 33 & 0.504 & 0.482 & 21 & 0.392 & 0.247 & 11 & $+$\\
\ciii \, ${\srm HWHM}$ & $\alpha_B$ & 0.039 & 0.046 & 56 & 0.282 & 0.312 & 30 & 0.084 & 0.171 & 17 & $+$\\
\ciii \, ${\srm HWHM}$ & $v_{\rm max}$ & 0.050 & 0.063 & 69 & 0.156 & 0.237 & 41 & 0.099 & 0.092 & 18 & $+$\\
\ciii \, ${\srm REW}$ & \feii \, $\lambda\,2070$ & 0.920 & 0.912 & 38 & 0.574 & 0.557 & 19 & 0.411 & 0.457 & 12 & $-$\\
\ciii \, ${\srm REW}$ & \feii \, $\lambda\,2400$ & 0.943 & 0.997 & 30 & 0.207 & 0.193 & 19 & 0.025 & 0.038 & 10 & $-$\\
\ciii \, ${\srm REW}$ & $\alpha_B$ & 0.880 & 0.967 & 38 & 0.674 & 0.881 & 19 & 0.681 & 0.527 & 12 & $+$\\
\ciii \, ${\srm REW}$ & $v_{\rm max}$ & 0.125 & 0.117 & 38 & 0.379 & 0.501 & 19 & 0.726 & 0.678 & 12 & $-$\\
\feii \, $\lambda\,2070$ & \feii \, $\lambda\,2400$ & {\bf 0.000} & {\bf 0.000} & 35 & 0.040 & 0.026 & 21 & 0.057 & 0.079 & 13 &
$+$ \\
\feii \, $\lambda\,2070$ & $\alpha_B$ & 0.133 & 0.110 & 43 & 0.763 & 0.771 & 21 & 0.347 & 0.226 & 15 & $+$\\
\feii \, $\lambda\,2070$ & $v_{max}$ & {\bf 0.005} & {\bf 0.003} & 43 & 0.414 & 0.383 & 21 & 0.096 & 0.080 & 15 & $+$ \\
\feii \, $\lambda\,2400$ & $\alpha_B$ & 0.048 & 0.079 & 35 & 0.116 & 0.167 & 21 & 0.581 & 0.648 & 13 & $+$\\
\feii \, $\lambda\,2400$ & $v_{\rm max}$ & {\bf 0.001} & {\bf 0.000} & 35 & 0.057 & 0.038 & 21 & 0.570 & 0.406 & 13 & $+$ \\
$\alpha_B$ & $v_{\rm max}$ & {\bf 0.001} &  {\bf 0.001} & 66 & 0.280 & 0.226 & 32 & {\bf 0.008} & {\bf 0.005} & 24 & $+$\\
\hline
\end{tabular}
\label{table:stats}
\end{table*}

The results of these correlation tests are given in Table
\ref{table:stats}.  The first two columns list the two indices we test.  
The next two columns give the
probabilities $P_{\tau}$ and $P_{r_s}$ to find a correlation by chance
between two uncorrelated quantities.  The column labeled $n$ is the number of
objects involved in the analysis.  Since we search for correlations
among 11 quantities, a total of 55 correlations tests are performed.
Since the number of objects is rather large for most pairs of indices,
we may reasonably consider as significant only those correlations with $P \leq 0.02$. 
With 55 correlations tested, we expect to detect one false correlation with such a 
probability, assuming they are independent.  In general, an excellent agreement is 
found between the results of the two statistical tests. The correlations
have also been investigated for the HIBAL and LIBAL QSO subsamples separately
and the results are given in Table \ref{table:stats}.  A few
correlations are detected only when considering the whole BAL QSO
sample, illustrating the importance of the size of the sample.

Among all significant correlations reported in Table
\ref{table:stats}, several link in an obvious manner quantities either
related by their definition or by a trivial common physical origin.
In the first category, as an example, the correlation between BI
and $v_{\rm max}$ simply reflects the fact that the wider the trough
the larger both BI and $v_{\rm max}$ (with BI $\leq v_{\rm max} -
5000$ according to their definitions).  Another such trivial
correlation is DI - $v_{\rm max}$: the most detached absorptions have
necessarily large $v_{\rm max}$.  In the second category, we may quote 
the correlations between \civ\, HWHM and \ciii \, HWHM and between 
the \feii\ indices.  Finally, a
few correlations in Table \ref{table:stats} are due to the fact that
two quantities are correlated with a third one : $v_{\rm max}$ is
correlated with \feii \, and $\alpha_B$ because, as we will see below,
the balnicity index is correlated with these quantities.  With all
these arguments in mind, we are left with six significant non-trivial
correlations : $p_0$ $-$ DI, BI $-$ \feii \, $\lambda$ 2400, BI $-$ 
$\alpha_B$, DI $-$ \civ \, HREW, $p_0$ $-$ \feii \, $\lambda$ 2070 and 
\civ \, HWHM $-$ $v_{\rm max}$.  They are all discussed in detail below and the most
interesting ones are illustrated in Fig. \ref{fig:corr}.  We have checked
that these correlations remain significant when we exclude the objects with 
the most extreme properties (e.g. the two LIBAL QSOs with DI $\geq 10$ in Fig. 
\ref{fig:corr}a).  

The degree of optical polarization, $p_0$, is essentially uncorrelated
with most indices apart from a few exceptions.  The most striking one is the 
anti-correlation between $p_0$ and the detachment index DI, already detected in 
Paper I and first suggested by Goodrich (\cite{GOO97}).  It is confirmed here with a higher 
significance ($P \simeq 0.002$) and on the basis of a larger sample. It means that, 
in average, BAL QSOs with \civ\, P Cygni profile \footnote{DI is small in P Cygni type
profiles since the absorption starts near zero velocity} are more polarized 
than objects with detached absorption troughs.  As can be inferred from Table 
\ref{table:stats}, this anti-correlation is strong for the whole BAL and the LIBAL 
QSO samples, but not detected in the HIBAL QSO sample. Fig. \ref{fig:corr} shows that 
the relation between $p_0$ and DI looks more like an inequality than a
real anti-correlation.  Indeed, for small values of DI, $p_0$ varies within
a large range of values (from 0 to 5\,\%), while for large values of DI, $p_0$ is
smaller.  This correlation indicates that DI may play a key-role in understanding 
the BAL QSO physics, as suspected by Hartig \& Baldwin (\cite{HAR86}). Since it is the most
important correlation involving polarization, it will be discussed in detail in Sect. 6.

The correlation between $p_0$ and \feii \,$\lambda$ 2070 is more
doubtful because it is significant only for the HIBAL QSO subsample
(marginal for the whole BAL QSO sample).  Moreover, the fact that
$p_0$ increases with the iron index does not support the usual
interpretation of a dilution of $p_0$ by an unpolarized pseudo-continuum 
formed of a large number of metastable iron emission lines.  Finally,
we do not see any obvious reason why $p_0$ should correlate with \feii 
\,$\lambda$ 2070 and not with \feii \,$\lambda$ 2400.  For all these 
reasons, we believe that this correlation is probably due to chance.

It is also important to remark that we do not detect any
correlation between $p_0$ and BI. Such a correlation was reported by
Schmidt \& Hines (\cite{SCH99}) in their Fig. 8 (although their correlation
was marginal with $P \sim$ \,0.05).  Neither do we recover the
correlation between $p_0$ and the slope of the continuum $\alpha_B$
discussed in Paper I and already interpreted as an artifact.

The correlations between BI and the \feii \,$\lambda$ 2400 index on
one hand and between DI and \civ \, HREW on the other hand have been
reported previously in W91.  They are confirmed here with larger
samples and with the use of BI averaged from the measurements of W91
and Korista et al. (\cite{KOR93}). The correlation between BI and the \feii
\,$\lambda$ 2400 index remains strong within the HIBAL QSO subsample
while it disappears in the LIBAL QSOs subsample. These two correlations 
are illustrated in Fig. \ref{fig:corr}.  
Another significant correlation involving BI is the correlation between BI 
and the slope of the continuum $\alpha_B$ (see Fig. \ref{fig:corr}). It 
was found in Paper I and is confirmed here. It indicates that BAL QSOs
with redder spectra display on average stronger absorption lines. Since
these two correlations involving BI appear to be important in the
description of the BAL phenomenon, they will be discussed later in the 
framework of the BALR models (Sect. 6.4).

The anti-correlation between DI and \civ \, HREW may possibly be
explained in terms of radiative pumping of the BEL photons.  Indeed,
for P Cygni profiles (DI $\ll$), the BAL trough cuts off the blue wing
of the \civ \, broad emission line (Turnshek \cite{TUR88}).  Photons are
absorbed and partly re-emitted in the red wing, therefore increasing the
equivalent width of the red part of the \civ \, emission.

Finally, the correlation between \civ \, HWHM and $v_{\rm max}$ $-$definitely 
less significant$-$ is not obvious to interpret. Although it may be spurious, 
it may also suggest that the dynamics of the BAL wind is connected 
in some way to the dynamics of the broad emission line region from which
it could be launched.

\section{Principal Component Analysis}

When we are dealing with a large number $n$ of BAL QSOs and a
large number $p$ of distinct parameters characterizing the QSO
properties, the study of the correlations two by two may be quite
complex.  Indeed, as can be seen from Table \ref{table:stats}, several
indices appear correlated, which makes it difficult to
disentangle the leading correlations and consequently the physical
mechanisms behind them.  The Principal Component Analysis (PCA) is an
alternative mathematical tool specifically designed to deal with such
multivariate problems.  Basically, the PCA identifies those parameters
that correlate together and decreases the complexity of the
problem by gathering them into only a few new parameters. In a
$p$-dimensional space, PCA identifies those directions in which the
cloud of points representing the QSOs is most elongated and uses them
to define a new system of coordinates.  Each axis in this new system is
called an eigenvector and is simply a linear combination of the
initial parameters. The first eigenvectors then contains most of the
variance observed in the sample and identify the most significant
correlations. For a complete mathematical description of multivariate
statistics, see Murtagh \& Heck (\cite{MUR87}).  For nice illustrations of the
application of PCA to some AGN data samples, see Boroson \& Green
(\cite{BOR92}), Francis \& Wills (\cite{FRA99}) and Boroson (\cite{BOR02}).

\begin{table}[htb]
\centering
\caption[ ]{Results of the Principal Component Analysis.
The first three principal components out of a total of 8 are given in
order of their contribution to the total variance.  The first row
gives the variances of the data along the direction of the
corresponding principal component.  The sum of all the variances is
equal to 8, the number of variables considered in the analysis.  The
variance of each principal component is given as ``proportion'' in the
second row and as ``the cumulative proportion'' in the third one.  
The rest of the table gives the correlation coefficients between the
original variables and the principal components.  Significant 
correlations are boldfaced, i.e. those ones ith
coefficients whose absolute values are \ $> 1/ \sqrt 8$ = 0.3536}
\begin{tabular}{lrrr}
\hline \\[-0.10in]
 & PC1 & PC2 & PC3 \\
\hline \\
Eigenvalue & 2.63 & 1.92 & 1.27 \\
Proportion & 32.9 & 24.1 & 15.9 \\
Cumulative & 32.9 & 57.0 & 72.9 \\
[0.10in]
Variable &  PC1 & PC2 & PC3 \\
\hline \\
$p_0$ & $-$0.1771 & {\bf 0.4795} & $-${\bf 0.4297} \\
BI & $-${\bf 0.5065} & 0.2048 & $-$0.1247 \\
DI & $-$0.2440 & $-${\bf 0.5507} & $-$0.1789 \\
$\alpha_{\rm B}$& $-${\bf 0.3867} & 0.0654 & 0.1423 \\ 
\civ \, ${\srm HREW}$ & $-$0.0014 & {\bf 0.5660} & $-$0.1301 \\
\civ \, ${\srm HWHM}$ & $-$0.1509 & 0.2636 & {\bf 0.7218} \\
\feii \, 2400 & $-${\bf 0.4865} & $-$0.1403 & $-$0.3062 \\
$v_{\rm max}$ & $-${\bf 0.4935} & $-$0.1053 & 0.3406 \\
\hline \\
\end{tabular}
\label{table:pca}
\end{table}

Since some objects have different wavelength coverages and/or their
spectra are of poor quality, there are obviously missing data for many
QSOs of our sample. Because a full data set is needed to run the
PCA, the number of useful objects will then depend on the 
indices taken into account. As a rule of thumb, the more numerous the
indices, the smaller the sample.  Moreover, we wish to avoid most of
the obvious correlations between closely related variables which would
be dominant in the first eigenvectors. After several trials, we made a
compromise between the size of the sample and the number of the
adopted parameters.  Together with $p_0$, all the parameters
describing the emission and the absorption properties of the \civ \,
line profile (BI, DI, \civ\,HREW, \civ\,HWHM and $v_{\rm max}$) 
are considered. Indeed, \civ\, measurements are available from medium
resolution spectra for a large number of QSOs of our sample (W91,
Korista et al. \cite{KOR93}).  We also consider one of the \feii\ intensity
indices : \feii \,$\lambda 2400$. The \feii\,$\lambda 2400$ index 
is favored since it is much stronger than the $\lambda 2070$ index.  
Finally, we add the slope of the continuum, $\alpha_B$, another important 
quantity in the description of BAL QSO spectral properties.  The indices 
related to \ciii \, are not considered because of the well-known blend with
\aliii \, and a possible blend with \feiii \, in some LIBAL QSOs (Hartig
\& Baldwin \cite{HAR86}).  Several PCA tests were performed including the
absolute B magnitude of the objects.  Since this quantity does
not correlate with any of the indices, it may be safely discarded from the 
analysis. However, we cannot definitely rule out any correlation with 
the intrinsic luminosity since some BAL QSOs may be significantly reddened
due to extinction (Sprayberry \& Foltz \cite{SPR92}, Reichard et al. \cite{REI03}). 
The final sample then consists in 30 BAL QSOs with good quality measurements for 
the 8 variables previously described.

The PCA was carried out with the Fortran code written by Murtagh \&
Heck (\cite{MUR87}).  Since the ranges of the variables are very different, the
original data are first centred and reduced to unit standard deviation
(correlation matrix option). 

We present the results in Table \ref{table:pca} which lists the three
most significant principal components (PC) and their projections upon
the eight variables described above.  Only the PCs with a variance
(eigenvalue) larger than 1 are considered.  More than half the total
variance of the sample (57\,\%) is accounted for by the first
two PCs which are elongated with variances larger than or of the order
of twice that of the original data.  We concentrate therefore
exclusively on these two principal components which we call PC1 and
PC2.
 
Apart from the usual redundant correlations, PC1 is dominated by the
correlation between the balnicity index BI and the strength of the
\feii \, emission  (and marginally $\alpha_B$), while PC2 correlates 
essentially $p_0$, DI and \civ \, HREW.  This analysis is in good agreement 
with the classical correlation tests performed in Sect. 3.  Indeed, two of the four
important correlations discussed in Sect. 3 ($p_0$ $-$ DI and DI $-$ \civ\, 
HREW) dominate the eigenvector PC2 while a third one (BI $-$ 
\feii\,$\lambda\,2400$) dominates PC1. Small differences with the previous
analysis (i.e. the significance of some correlations) may be explained 
in terms of the smaller sample considered here. Therefore, the main result 
emerging from the PCA is that BI, \feii \, $\lambda 2400$ and maybe $\alpha_B$ 
on one hand, and $p_0$, DI and \civ \, HREW, on the other hand,  
vary together.  As will be discussed in Sect. 6, PC1 could be connected to 
the accretion rate of matter (since this interpretation may qualitatively 
reproduce the correlation BI - \feii), while PC2 could be related to the 
orientation of the BAL QSO with respect to the line of sight.  The existence 
of two PCs outlines that, although important, orientation is probably not the 
only parameter governing the BAL QSO properties.

\begin{figure}[t]
\resizebox{\hsize}{!}{\rotatebox{0}{\includegraphics{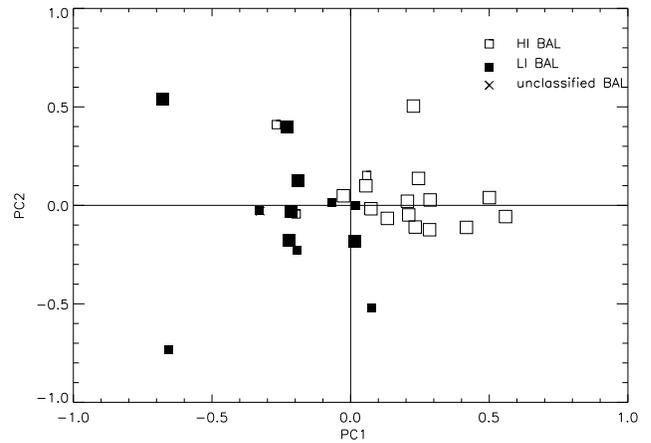}}}
\caption[]{Distribution of the 30 BAL QSOs with respect to PC1 and
PC2.  Open squares represent HI BAL QSOs, filled squares LI BAL QSOs
while the cross represents the only unclassified BAL QSO considered in
the PCA. Large symbols refer to bright QSOs with $M_B \leq -28$ and
small symbols represent fainter QSOs with $M_B > -28$.}
\label{fig:pca0}
\end{figure}

Fig. \ref{fig:pca0} illustrates the distribution of the 30 BAL QSOs in
the PC1-PC2 space. Different symbols have been used to separate the
HIBAL and LIBAL QSO subsamples and to distinguish the lower luminosity 
objects (with $M_B > -28$)
from the brighter ones (with $M_B \leq -28$). The LIBAL QSO subsample is
essentially located in the $\rm{PC1} \le 0$ region while most of the
HIBAL QSOs have ${\rm PC1} > 0$.  The HIBAL and LIBAL QSO subsamples 
are therefore better discriminated with PC1 than with PC2.  Indeed, 
the LIBAL QSOs have the largest BI and \feii \, emission or at least 
they display the most extreme values (see Fig. \ref{fig:corr}). If PC2
is indeed related to orientation, it plays a less important role in 
explaining the differences observed between HIBAL and LIBAL QSOs. 
Among the LIBAL QSOs, there is also a trend (not represented in Fig.
\ref{fig:pca0}) that the objects with stronger low ionization troughs 
have more negative values of PC1, i.e. the S LIBAL QSOs have more 
negative PC1 than W or M LIBAL QSOs, the latter having values of PC1 closer
to those of HIBAL QSOs. Finally, interestingly enough,
fainter QSOs appear to be located in the ${\rm PC1} \leq 0 $ 
part of the diagram.

\section{BAL QSOs spectropolarimetry}

\subsection{Description of the observed sample}

Seven BAL QSOs were chosen for spectropolarimetry because of their brightness,
their high degree of linear polarization in the V filter (Paper I) and 
their redshift adequate for the observation of the \civ \, absorption line. 
Moreover, all these objects are LIBAL QSOs with strong, weak or marginal 
absorptions from low ionization species.  They also span a range of other properties :
\object{B2225$-$0534} and \object{B2240$-$3702} are BAL QSOs with P~Cygni \civ \, 
profiles (DI $\ll$) while \object{B1246$-$0542} and \object{B1011$+$0906} are objects with 
detached \civ \, troughs (DI $\gg$).  \object{B1413$+$1143} is a gravitationally lensed
QSO with evidence for a micro-lensing effect (Angonin et al. \cite{ANG90}, Hutsem\'ekers 
\cite{HUT93}) and \object{B0059$-$2735} is the prototype of the so-called ``Iron Lo BAL QSOs''
(Hazard et al. \cite{HAZ87}, Becker et al. \cite{BEC97}).  Spectropolarimetry of 
\object{B0059$-$2735} has been published elsewhere (Lamy \& Hutsem\'ekers \cite{LAM00b}) and 
will not be discussed again.

Spectropolarimetry of \object{B2225$-$0534} has been abundantly described in the
literature in the last two decades (Stockman et al. \cite{STO81}, Cohen et
al. \cite{COH95}, Goodrich \& Miller \cite{GOO95}, Ogle et al. \cite{OGL99}). 
Since there is no indication of variability in this object, these data have enabled us
to check the whole observation/reduction procedure.

\begin{figure*}
\resizebox{\hsize}{!}{\rotatebox{0}{\includegraphics{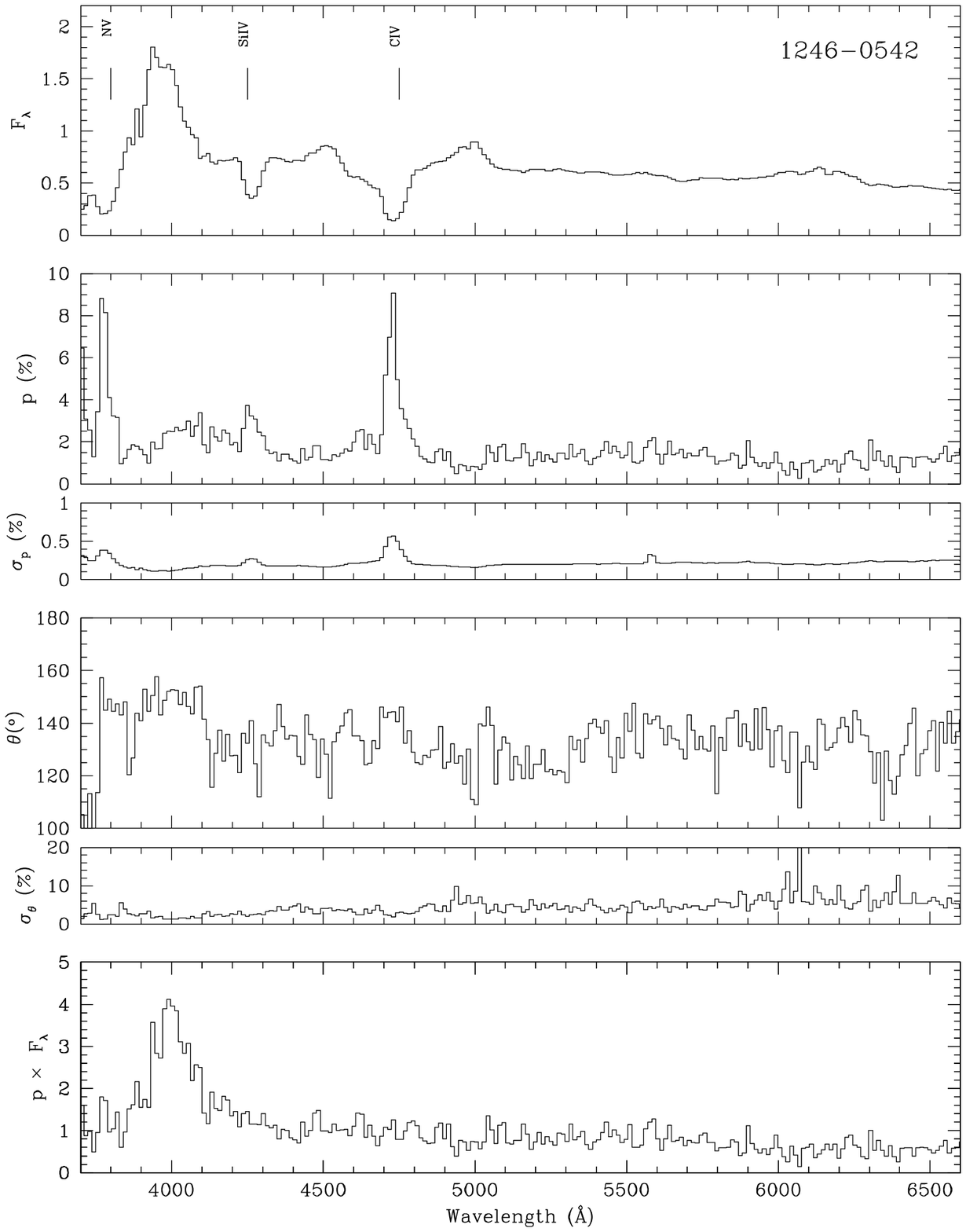}}
{\includegraphics{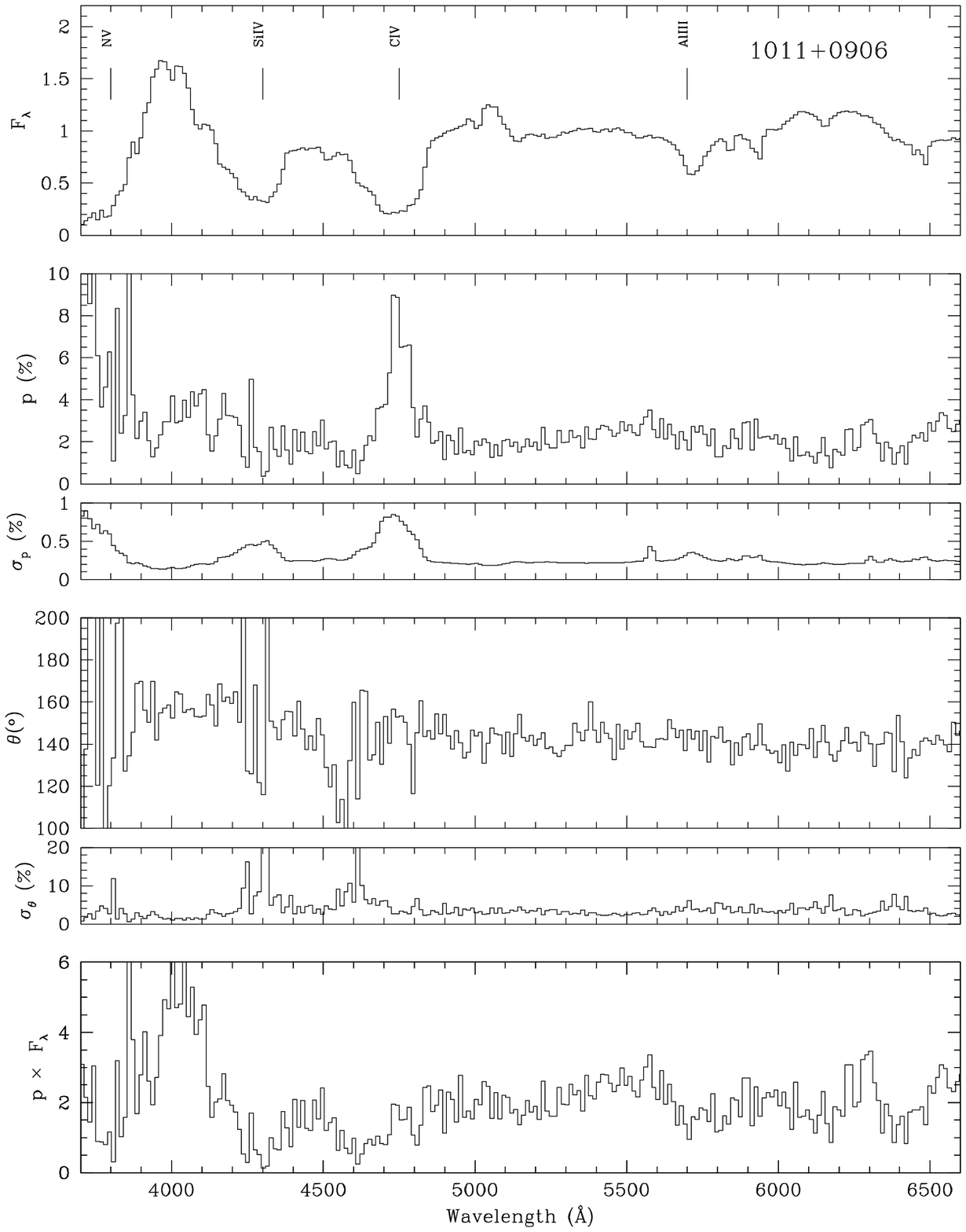}}} \\ 
\resizebox{\hsize}{!}{\rotatebox{0}{\includegraphics{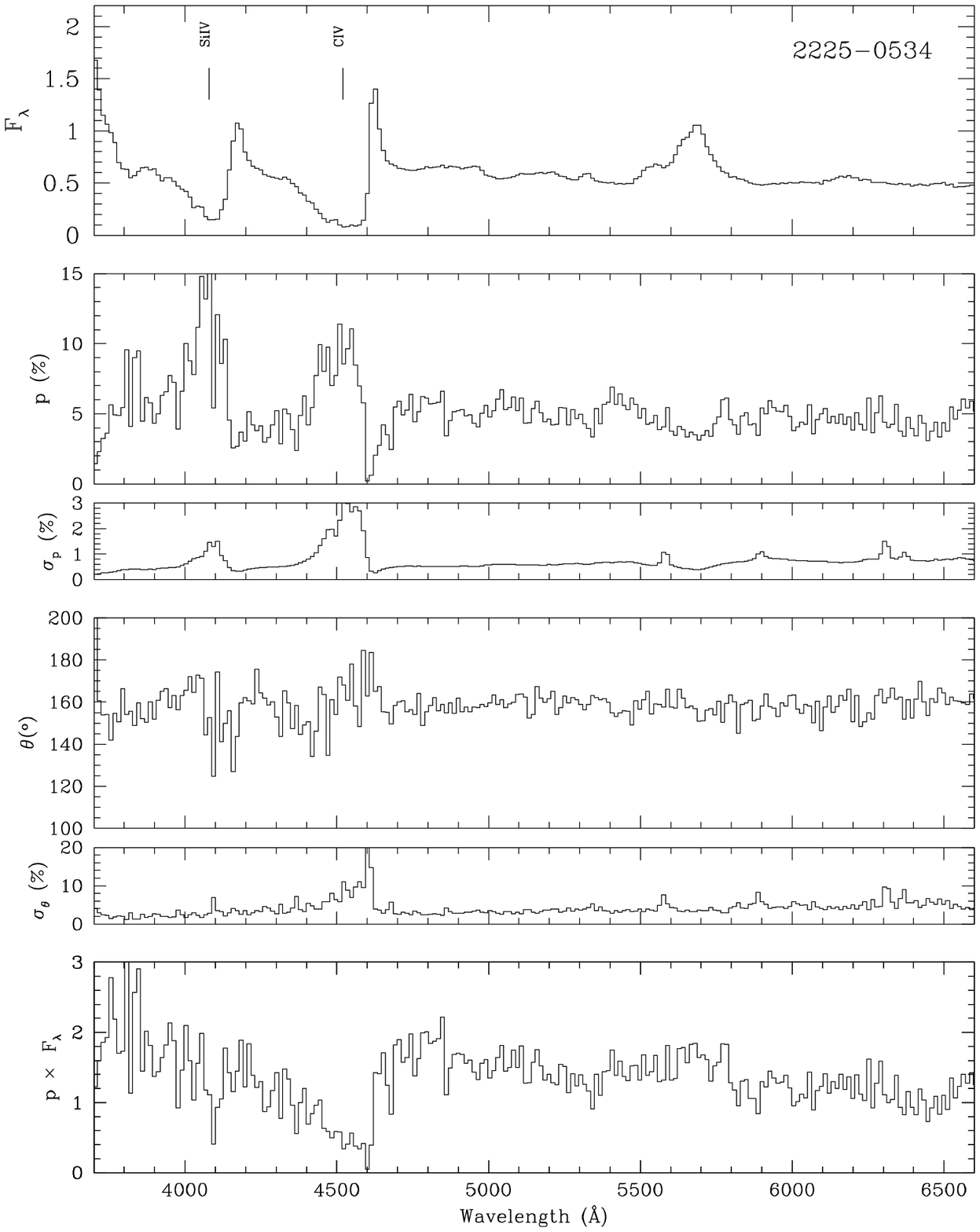}}
{\includegraphics{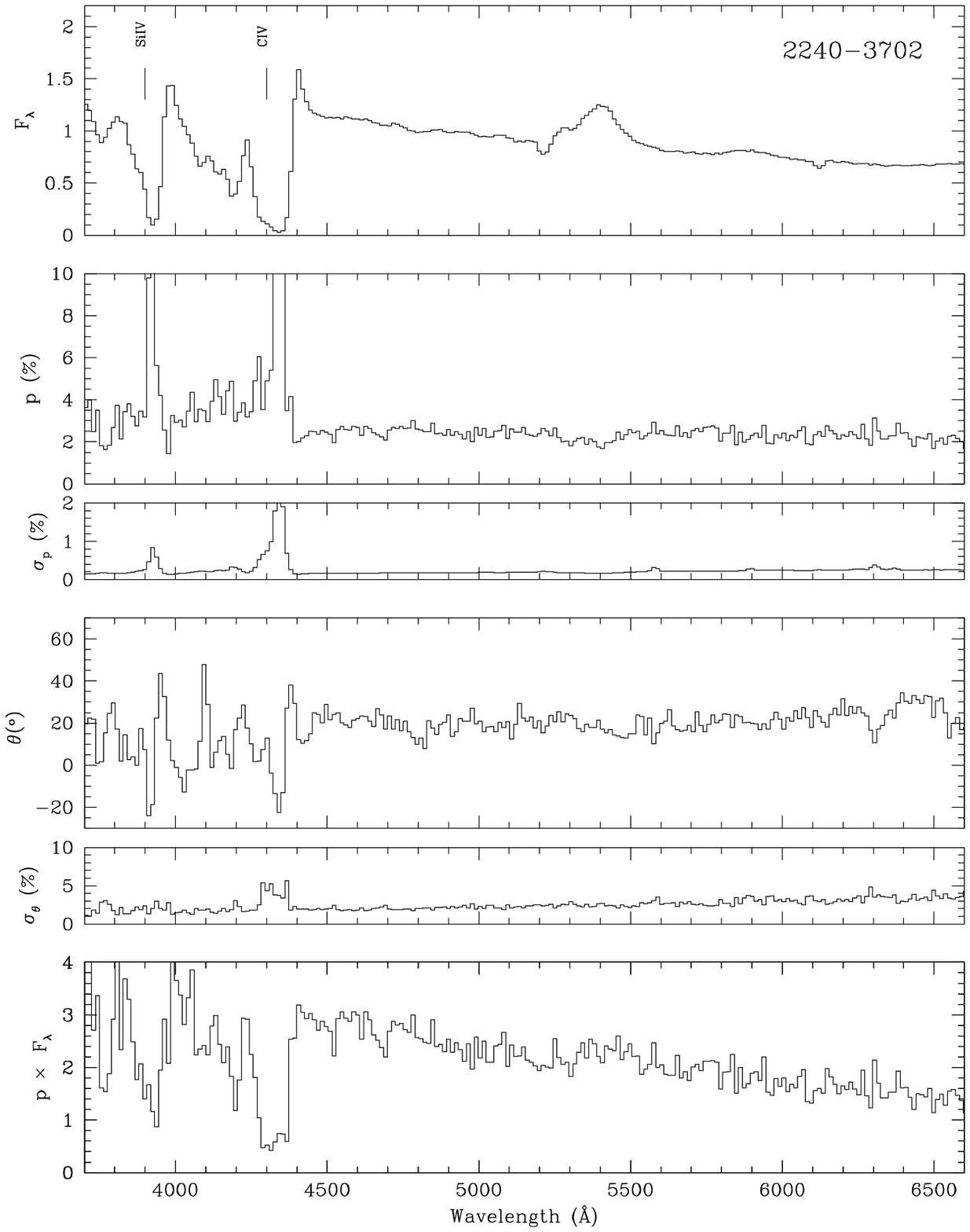}}}
\caption[]{Spectropolarimetric data for the BAL QSOs \object{B1246$-$0542} (top
left), \object{B1011$+$0906} (top right), \object{B2225$-$0534} (bottom left) and
\object{B2240$-$3702} (bottom right). For each object, there are six
panels; from top to bottom : total flux $F_{\lambda}$, degree of
polarization $p$, associated error $\sigma_p$, polarization position
angle $\theta$, associated error $\sigma_{\theta}$, and polarized flux
$p\,\times\,F_{\lambda}$.  $p$ has not been debiased. The main
absorption lines are indicated.}
\label{fig:spectropola1}
\end{figure*}

\begin{figure*}
\resizebox{\hsize}{!}{\rotatebox{0}{\includegraphics{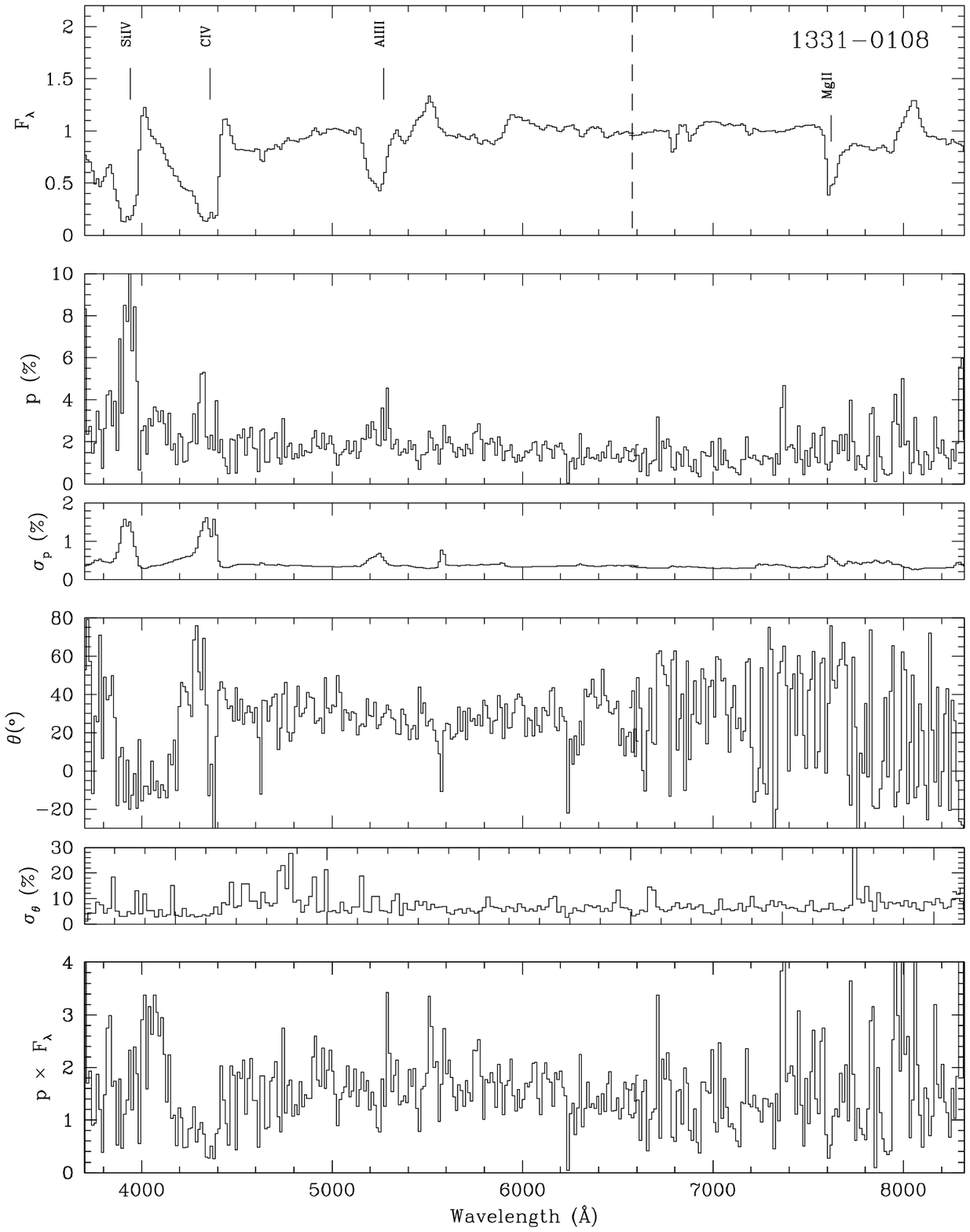}}
{\includegraphics{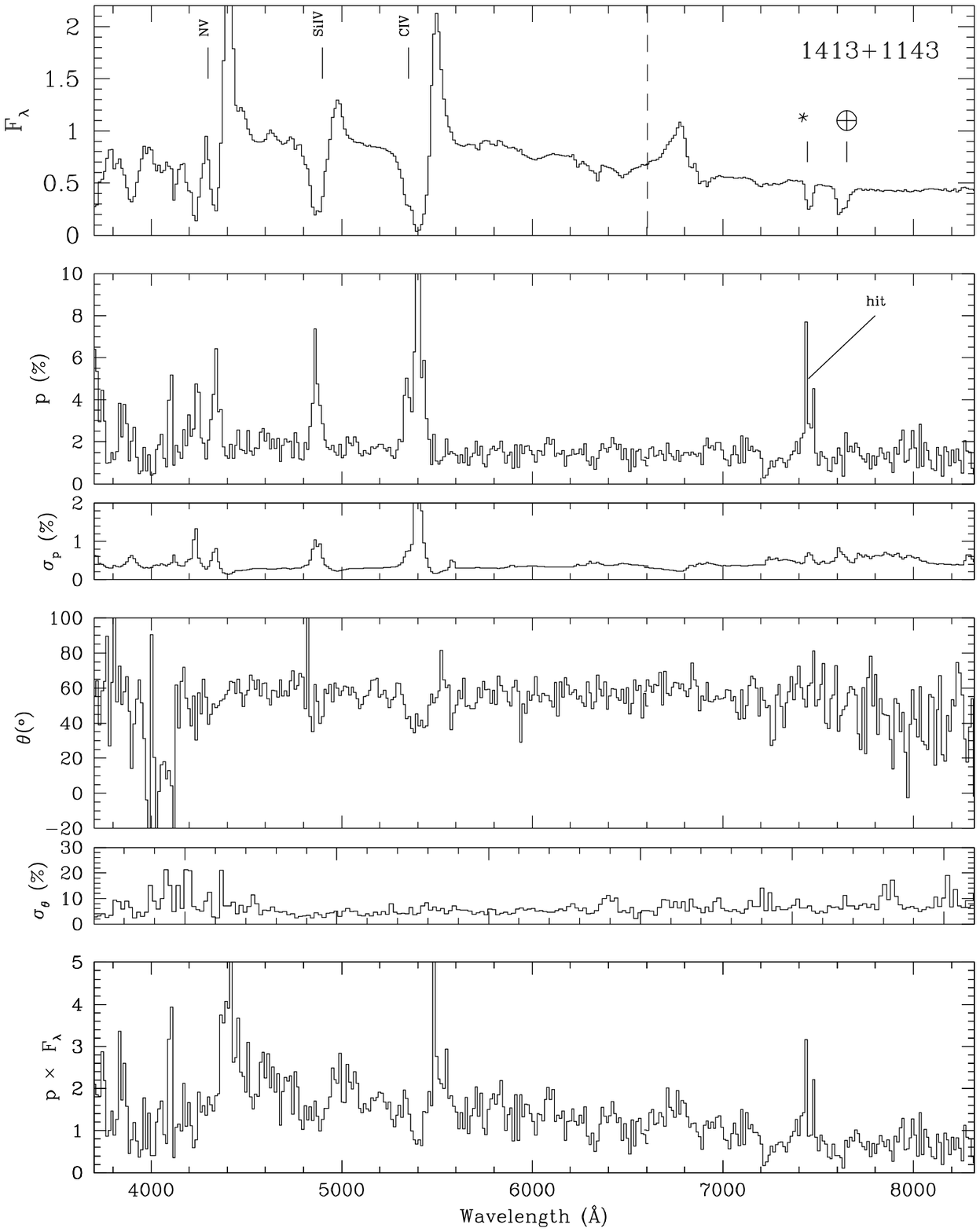}}} 
\caption[]{Spectropolarimetric data for the BAL QSOs \object{B1331$-$0108}
(left) and \object{B1413$+$1143} (right). Panels are as in
Fig.~\ref{fig:spectropola1}.  The B300 and R300 spectra were cut on
their reddest and bluest parts respectively and merged together at the
wavelength indicated by the vertical dashed line.  The main absorption
lines are indicated as well as features due to cosmic rays hits. The
asterisk in the spectrum of \object{B1413$+$1143} indicates an intervening
absorption system at $z \sim 1.66$.  The Earth symbol indicates a
telluric absorption.}
\label{fig:spectropola2}
\end{figure*}

\subsection{Observations and data reduction}

The spectropolarimetric observations were carried out at the 3.6m
telescope (La Silla, ESO, Chile) equipped with EFOSC1 during two runs
in April 5--8, 1995 and September 10--12, 1996.

Spectropolarimetry is performed with EFOSC1 by inserting a Wollaston
prism in the filter wheel and a grism in the grism wheel (di Serego
Alighieri \cite{DIS98}). The Wollaston prism splits the incoming radiation
into two orthogonally polarized spectra separated by 20$\farcs$ Each
object positioned on the slit has two spectra projected on the CCD
detector. The measurement of their ratio at two different Wollaston
angle positions separated by 45$^{\circ}$ gives the two Stokes
parameters describing the polarization (no half-wave plate was
available at that time).  In practice, the Wollaston prism does not
move but rotates together with the whole EFOSC1 instrument, including
the slit.  In our case, this is not critical since our objects are
point-like.  In order to avoid overlapping of the object and sky
spectra, a mask is placed in front of the slit; it is
transparent only in slots whose length is equal to the beam separation
produced by the Wollaston prism.  The 2D image recorded on the CCD
then consists of a number of alternate bands, two of them containing
the two orthogonally polarized spectra of the object, the others being
devoted to the sky.

The detector was a 512 $\times$ 512 Tek CCD (ESO\#26) with an inverse
gain of 3.8 electrons per ADU and a read-out-noise of 7.9 electrons.
The pixel size was 27 $\mu$m corresponding to 0$\farcs$605 on the sky
(Melnick et al. \cite{MEL89}).  The slit had a width of 2$\farcs$ The seeing
was typically around 1$\farcs$4.  The B300 grism was used for the seven
objects giving a spectral coverage of 3600-6800 \AA.  For two objects,
the R300 grism was also used to enlarge the useful spectral range up 
to $\sim 8300$ \AA \, allowing us to observe the \mgii \,BAL. The 
spectral resolution was 6.7 \AA \, per pixel.  For each grism and position 
angle of the Wollaston, three frames were secured for a total exposure time of 
90 minutes per Wollaston position. The splitting in shorter
exposures ensures that the contamination by cosmic rays can be corrected.

Data reduction was accomplished with procedures developed within the
ESO MIDAS package. The 2D raw data are first corrected for bias and
flat-field.  Cosmic ray events are interactively corrected and
replaced by the median value of a small surrounding area.  However, it
is sometimes impossible to correct for some cosmic rays falling on
the object spectra.  These cosmic rays are easily detected by
comparing the individual frames and they are flagged in the final 1D 
spectra with a 'hit' label.  The two object spectra and the four sky spectra 
(two on both sides of the object) are individually extracted with exactly 
the same procedure.  The six spectra are calibrated in wavelength using 
He+Ar internal lamp spectra.  Since we observed small offsets
between corresponding He+Ar emission lines in orthogonally polarized
spectra, six individual wavelength calibration curves were produced,
and each object/sky spectra was independently wavelength calibrated.
The sky spectra are averaged on both sides of the object and
subtracted from the object spectra.  Spectra are then corrected for
the atmospheric extinction.  Since there was no HWP plate in the setup, 
a flux calibration was done in order to correct for a possible difference
of transmission of the Wollaston prism + grism for the ordinary and
extraordinary rays.  For that purpose, we observed spectrophotometric 
standard stars, assumed to be unpolarized, with the same instrumental setup.  
The standard star \object{EG274} (Stone \& Baldwin \cite{STO83}, Baldwin \& Stone 
\cite{BAL84}) was found to be very stable from night to night and was used to obtain the 
individual flux calibration curves. For that star, we measure a weak polarization 
degree ($p \leq 0.3 \%$), more or less constant with wavelength, and in 
agreement with the observations by di Serego Alighieri (\cite{DIS94}) with the same 
instrumental setup.  Although the instrumental polarization induced by the instrument 
is therefore small, we took account of it by performing the flux calibration.  
Finally, it was necessary to rebin the data in wavelength in order to achieve a 
polarization uncertainty of $\sim$ 0.2 - 0.3\,\% in the continuum.  The spectra 
are rebinned on two original pixels, which corresponds to a spectral resolution of 
about 13 \AA.  The associated errors are calculated by propagating the errors from 
the photon noise in the object and sky spectra.   

The Stokes parameters $q$ and $u$, the degree of polarization $p$ and the
angle of polarization $\theta$ are computed from these rebinned
spectra in the usual way (di Serego Alighieri \cite{DIS98}).  Rebinning
is done before the calculation of the Stokes parameters since $q$ and $u$
have non-gaussian distribution errors (Clarke \& Stewart \cite{CLA86}). The
photon noise is propagated throughout the equations in order to
obtain $\sigma_p$ and $\sigma_{\theta}$ in every bin of wavelength.
Although $p$ is a well-known biased quantity (Simmons \& Stewart
\cite{SIM85}), we do not de-bias it since our data have large S/N
ratios. Finally, we calculate the polarized flux by multiplying the
total flux by the degree of polarization. All these results are given in Figs. 
\ref{fig:spectropola1} and \ref{fig:spectropola2} for the six BAL QSOs . 

A polarimetric standard star, \object{HD161291}, was observed during both runs with the 
same setup.  The interstellar polarization wavelength dependence was found in good 
agreement with that measured by di Serego Alighieri et al. (\cite{DIS94}) using the same 
setup, and with the parameters given by Serkowski et al. (\cite{SER75}).

\subsection{Description of individual objects}

The spectropolarimetry of individual BAL QSOs is discussed in this
section.  We emphasize mainly their differences as well as 
new results that were not previously reported by Schmidt \& Hines
(\cite{SCH99}) and more particularly by Ogle et al. (\cite{OGL99}) in their
spectropolarimetric atlas.  Note that Ogle et al. (\cite{OGL99}) use $q'$ 
instead of $p$. $q'$ is the Stokes parameter $q$ rotated by the angle
$2\,\overline{\theta}$ where $\overline{\theta}$ is the mean
continuum polarization angle.  Although $q'$ is unbiased, it has the
disavantadge that its value is difficult to interpret when measured in 
emission or absorption lines with significant rotation of $\theta$.

1. \object{B1246$-$0542} $--$ This BAL QSO has deep broad absorption lines
clearly detached from the corresponding emission lines.  The
polarization rises strongly in the \civ \, and \siiv \, broad
absorption lines, mainly in the deepest troughs, while it slightly
decreases in the emission lines.  There is no rotation of the
polarization accross those lines.  The polarized flux of
\object{B1246$-$0542} is remarkable: first, it is nearly featureless with no
trace of absorption, which is quite unusual among BAL QSOs (Schmidt
\& Hines \cite{SCH99}, Ogle et al. \cite{OGL99}). Second, it shows an unusually
large \nv\ emission, with no other emission lines.  Apparently
unnoticed, such a large \nv\ emission is in fact observed in the
polarized flux of several BAL QSOs from the spectropolarimetric
atlas of Ogle (\cite{OGL99}), as well as in \object{B1011$+$0906}.  A possible
interpretation of this higher \nv\ emission in the polarized flux, 
in contrast with other lines such as \civ\ and \siiv , could be an 
extra \nv\ emission due to resonance scattering of Ly$\alpha$ photons 
(Surdej \& Hutsem\'ekers \cite{SUR87}), polarized due to a non-spherical
geometry of the BALR (Lee \& Blandford \cite{LEE97}).

2. \object{B1011$+$0906} $--$
The spectrum of this BAL QSO has some similarities with \object{B1246$-$0542}, 
with deep absorptions well detached from relatively faint
emission lines, and a strong \nv\, emission.  As usual, the polarization 
increases within the \civ\, trough. The polarized flux is characterized 
by the presence of a large \nv \, emission. However, compared to \object{B1246$-$0542},
the polarized spectrum is complex and difficult to interpret.  In particular, 
the polarization does not increase in the \siiv\, 
trough while there are some absorption seen in the polarized flux close
to the location of the \siiv\, emission line.  This latter absorption 
possibly results from \civ\, absorption at velocities larger than 
$25000\,{\rm km\,s}^{-1}$ which could be superimposed on the \siiv\, emission 
line.  The \aliii\, BAL is polarized as the continuum.  The rotation of 
$\theta$ near 4600 \AA \, is not significant at the 2 $\sigma$ level, the 
uncertainty on $\theta$ in this region being large due to a decrease of $p$.  
However, if real, it may also be due to complex absorptions that are difficult 
to identify in this rather noisy part of the polarized flux.

3. \object{B2225$-$0534} $--$ This object is the prototype of the BAL QSOs 
with \civ\, P cygni type profiles.  The spectropolarimetry of \object{B2225$-$0534} has 
been published by Stockmann et al. (\cite{STO81}), Cohen et al. (\cite{COH95}), 
Goodrich \& Miller (\cite{GOO95}) and Ogle et al. (\cite{OGL99}). We confirm the 
main results found by these authors : the continuum is highly polarized with $p
\sim 5$\,\%, the broad emission lines are unpolarized with a large
drop of $p$ in the \civ \, emission line, and the polarization
strongly increases in the broad absorption lines.  The polarization
angle is constant throughout the spectral domain.  The emission lines
are absent from the polarized flux except a small residual from \ciii.
Since no indication of variability has been reported for this object
by previous authors, the fact that our results are in very good
agreement with theirs provides a useful cross-check of our reduction.

4. \object{B2240$-$3702} $--$ This BAL QSO is quite similar to \object{B2225$-$0534}.  
Indeed, it shows P Cygni type profiles and nearly black broad absorption lines.  But
\object{B2240$-$3702} is also different with very weak emission lines and a double 
trough structure in the \civ \, and \siiv \, absorptions.  The polarization 
strongly increases in the low velocity part of the absorption reaching a
maximum of $\sim$ 25 \% \ in \civ.  There is a dip of the polarization at the
position of the emission lines. The polarization redward of \civ \, is flat ($p
\sim$ 2 \,\%) with a small decrease at the location of \ciii. The polarization 
angle rotates accross the broad absorption lines and is constant elsewhere.  
The broad absorption lines appear in the polarized flux although shallower.
There is no evidence for any emission line in the polarized flux.

5. \object{B1331$-$0108} $--$ The data for this BAL QSO have a lower signal to
noise ratio.  There are however indications that the polarization
increases in the \civ, \siiv \, and \aliii \, absorption lines while
there is no clear evidence for a rise of $p$ in the \mgii \, trough.
Nothing can be said about the emission lines.

6. \object{B1413$+$1143}\footnote{Note that there is an error in the
polarization position angle data of \object{B1413$+$1143} displayed in Lamy \&
Hutsem\'ekers (\cite{LAM99})} $--$ This BAL QSO is a four component
gravitational mirage (Magain et al. \cite{MAG88}).  Again, the polarization
strongly increases in the absorption troughs.  Although the
absorptions of \civ \, and \siiv \, are nearly black in the total
flux, they are shallower in the polarized flux.  There is also a
rotation of the polarization angle accross the \civ \, absorption
line.  Nevertheless, the most outstanding result in this object is the
prominence of the emission lines in the polarized flux, indicating
that they are polarized at a similar level as the continuum.  Only the
semi-forbidden \ciii \, line is absent in the polarized flux.  Our
results are in good agreement with those published previously by
Goodrich \& Miller (\cite{GOO95}), Schmidt \& Hines (\cite{SCH99}) and Ogle et
al. (\cite{OGL99}), despite of the fact that the total flux and the
polarization of \object{B1413$+$1143} are known to be variable (e.g. Goodrich \& Miller 
\cite{GOO95}, {\O}stensen et al. \cite{OST97}).

\begin{table}[t]
\caption[ ]{The spectropolarimetric indices.  The measurements were done using
the spectropolarimetric data illustrated in Figs. \ref{fig:spectropola1} and
\ref{fig:spectropola2} and the polarization spectra of Ogle et al. (\cite{OGL99}).  
The continuum polarization, $p_0$, and its associated error, $\sigma_p$, are 
taken from Table 1. $p_0$, $\sigma_p$ and $p_{max}$ are in \%
while the other quantities are unitless.}
\begin{tabular}{lcrrrrrrr}
\hline \\[-0.10in]
Object & $p_0$ & $\sigma_p$ & $p_e/p_c$ &  $p_e/p_c$ & 
SI  & $p_{max}$\\
       &       &            &  \civ     &  \ciii     &  & 
\\[0.05in] \hline \\[-0.10in]
\object{B0019$+$0107}  &  0.98  &  0.02  &  0.90  &  1.00  &     -  &    4.8$\pm$0.6 \\  
\object{B0043$+$0048}  &  0.09  &  0.06  &     -  &     -  &     -  &    4.4$\pm$0.9 \\ 
\object{B0059$-$2735}  &  1.49  &  0.02  &     -  &     -  &     -  &   26.1$\pm$4.0 \\ 
\object{B0105$-$2634}  &  2.41  &  0.08  &  0.50  &  0.60  &  0.56  &   10.5$\pm$1.6 \\   
\object{B0137$-$0153}  &  1.09  &  0.05  &  0.50  &  0.50  &     -  &    6.5$\pm$2.1 \\  
\object{B0146$+$0142}  &  1.24  &  0.02  &  0.40  &  0.70  &  0.60  &    4.3$\pm$0.7 \\ 
\object{B0226$-$1024}  &  1.81  &  0.01  &  0.60  &  0.70  &  0.50  &    7.3$\pm$0.8 \\ 
\object{B0842$+$3431}  &  0.51  &  0.01  &  0.80  &  1.00  &  0.60  &    2.9$\pm$0.4 \\  
\object{B0903$+$1734}  &  0.67  &  0.02  &  0.30  &  0.70  &     -  &    5.2$\pm$1.3 \\ 
\object{B0932$+$5006}  &  1.11  &  0.02  &  0.70  &  0.80  &  0.85  &    2.7$\pm$0.7 \\ 
\object{B1011$+$0906}  &  2.00  &  0.08  &  0.78  &  0.66  &  0.74  &    9.0$\pm$0.8 \\ 
\object{B1212$+$1445}  &  1.49  &  0.03  &     -  &  0.60  &  1.00  &    4.3$\pm$1.0 \\  
\object{B1232$+$1325}  &  3.19  &  0.04  &  0.50  &  0.79  &  0.88  &   12.2$\pm$2.6 \\ 
\object{B1235$+$0857}  &  2.53  &  0.07  &     -  &  0.70  &  1.00  &    3.3$\pm$0.8 \\ 
\object{B1246$-$0542}  &  1.26  &  0.01  &  0.63  &  0.77  &  0.13  &    7.7$\pm$0.6 \\  
\object{B1331$-$0108}  &  1.55  &  0.14  &  0.70  &  0.30  &     -  &    7.2$\pm$1.4 \\  
\object{B1333$+$2840}  &  4.67  &  0.02  &  0.20  &  0.50  &  0.76  &    8.5$\pm$0.9 \\  
\object{B1413$+$1143}  &  1.52  &  0.04  &  0.90  &  1.00  &  0.66  &   13.2$\pm$2.2 \\  
\object{B1524$+$5147}  &  3.49  &  0.01  &  0.50  &  0.70  &  0.60  &    6.0$\pm$0.2 \\   
\object{B2225$-$0534}  &  4.26  &  0.02  &  0.21  &  0.70  &  0.93  &   10.7$\pm$2.9 \\  
\object{B2240$-$3702}  &  2.09  &  0.19  &  0.78  &  0.74  &  0.85  &   25.5$\pm$2.1 \\  
\hline  
\end{tabular}
\label{table:specmeas}
\end{table}

\subsection{Definition of spectropolarimetric indices}

\begin{table}[t]
\caption[ ]{Results of correlation tests between spectropolarimetric indices,
$p_0$ and three indices derived from optical spectra. $P_{\tau}$ and $P_{r_s}$ give the 
probabilities that the correlation between the two indices are due to chance.  
$n$ is the number of points considered for the correlation. A correlation is 
emphasized if $P \leq 0.05$.  The last column gives the sign of 
the correlation}
\begin{tabular}{llrrrc}
\hline \\[-0.10in]
Index 1 & Index 2 & $P_{\tau}$ & $P_{r_s}$ & $n$ & 
\\[0.05in] \hline \\[-0.10in]
$p_0$ & $p_e/p_c$ \civ & 0.067 & 0.091 & 17 & $-$\\
$p_0$ & $p_e/p_c$ \ciii & 0.147 & 0.167 & 19 & $-$\\
$p_0$ & SI & 0.224 & 0.254 & 15 & $+$\\
$p_0$ & $p_{\rm max}$ & {\bf 0.029} & {\bf 0.019} & 22 & $+$\\
$p_e/p_c$ \civ & $p_e/p_c$ \ciii & {\bf 0.012} & {\bf 0.018} & 17 & $+$\\
$p_e/p_c$ \civ & SI & 0.440 & 0.647 & 13 & $-$\\
$p_e/p_c$ \civ & $p_{\rm max}$ & 0.966 & 0.936 & 17 & $+$\\
$p_e/p_c$ \civ & BI & 0.523 & 0.583 & 17 & $-$\\
$p_e/p_c$ \civ & DI & 0.579 & 0.640 & 17 & $+$\\
$p_e/p_c$ \civ & $\alpha_B$ & 0.797 & 0.763 & 17 & $+$\\
$p_e/p_c$ \ciii & SI & 0.553 & 0.714 & 15 & $-$\\
$p_e/p_c$ \ciii & $p_{\rm max}$ & 0.823 & 0.777 & 19 & $-$\\
$p_e/p_c$ \ciii & BI & 0.970 & 0.951 & 19 & $+$\\
$p_e/p_c$ \ciii & DI & 0.852 & 0.913 & 19 & $+$\\
$p_e/p_c$ \ciii & $\alpha_B$ & 0.823 & 0.775 & 19 & $-$\\
SI & $p_{\rm max}$ & 0.959 & 0.947 & 15 & $-$\\
SI & BI & 0.919 & 0.914 & 15 & $+$\\
SI & DI & {\bf 0.033} & 0.073 & 15 & $-$\\
SI & $\alpha_B$ & 0.203 & 0.221 & 15 & $+$\\
$p_{\rm max}$ & BI & {\bf 0.008} & {\bf 0.004} & 22 & $+$\\
$p_{\rm max}$ & DI & 0.080 & 0.093 & 22 & $-$\\
$p_{\rm max}$ & $\alpha_B$ & 0.089 & 0.059 & 22 & $+$\\
\hline  
\end{tabular}
\label{table:spec_res}
\end{table}

In this section we define four indices in order to describe the
polarization of the emission and absorption lines in BAL QSOs.  The
measurement of these indices has been carried out on our spectropolarimetric 
data and on the spectra from Ogle et al. (\cite{OGL99}).  They are given in Table 
\ref{table:specmeas}. For the five BAL QSOs of our sample with good quality
data and which were also observed by Ogle et al. (\cite{OGL99}) (i.e. 
\object{B0059$-$2735}, \object{B1011$+$0906}, \object{B1246$-$0542}, 
\object{B1413$+$1143} and \object{B2225$-$0534}), the measurements are 
done independently for both data sets. Since they are in good agreement within 
uncertainties, we only give average values in Table \ref{table:specmeas}.

The first two quantities measure the polarization in the \civ \, and
\ciii \, emission lines relative to the polarization of the redward
adjacent continuum. They are given as $p_e/p_c$ \civ \, and $p_e/p_c$
\ciii \, in Table \ref{table:specmeas}.  The other two indices are
related to the polarization in the \civ \, absorption line. The first
one is the maximal polarization in the \civ\ trough, $p_{\rm max}$.
The second one, denoted SI, is the ratio of the \civ \, absorption depths 
(measured at the wavelength of the deepest \civ \, absorption)
in the polarized flux $p \times F_{\lambda}$ and in the total flux 
$F_{\lambda}$ where the continuum has been normalized to unity ($F_c$ 
is the flux in the continuum and $p_c$ is the polarization of the continuum)
i.e., 
\begin{equation}
{\rm SI} = \frac{1-\frac{p\,\times\,F_{\lambda}}{p_c\,\times\,F_c}}
{1-\frac{F_{\lambda}}{F_c}} 
\end{equation}
Of course, we have $0 \leq {\rm SI} \leq 1$. SI is a measure of the strength of the
absorption in the polarized flux with respect to the absorption in the
direct flux.

The values given in Table \ref{table:specmeas} are measured for
21 objects with good signal to noise data.  The uncertainties on $p_{\rm max}$
are taken from Ogle et al. (\cite{OGL99}) or they have been measured from our data (Figs. 
\ref{fig:spectropola1} and \ref{fig:spectropola2}).  The uncertainties on the 
other indices are more difficult to estimate, namely since they depend on the 
position of the continuum and the signal to noise in the polarized spectrum.
Based on two independent sets of measurements, we estimate that the typical
uncertainties of these quantities are $\sim 10-15\,\%$.

\subsection{Correlation analysis}

Table \ref{table:spec_res} gives the results of the correlation
analysis performed between the four spectropolarimetric indices
described in the previous section, the broad band polarization $p_0$
and three important and well-measured spectroscopic indices (BI, DI and $\alpha_B$) 
given in Table 1.  Correlations have again been searched for using 
the Kendall $\tau$ and Spearman $r_s$ rank correlation coefficients.
With only 21 objects, we relax our criterion and consider as significant 
correlations with $P \leq 0.05$.  With 22 correlations (assumed to be
independent), we expect that one correlation with $P \leq 0.05$ is due 
to chance.

\begin{figure}
\resizebox{\hsize}{!}{\rotatebox{0}{\includegraphics{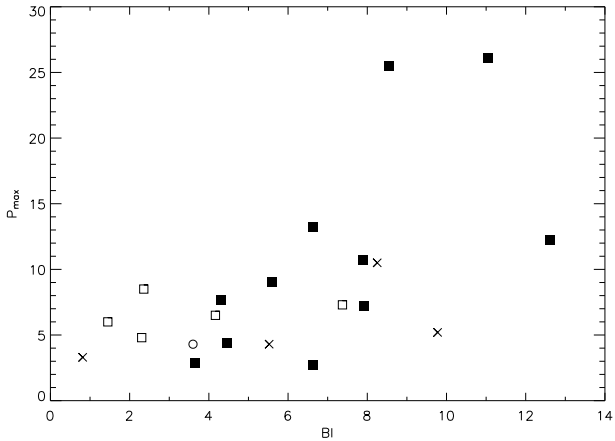}}} 
\caption[]{The correlation between the maximum value of the polarization in the \civ\,
absorption line and the balnicity index BI.  Open squares are HIBAL QSOs, filled squares 
are LIBAL QSOs and crosses are unclassified BAL QSOs. The open circle corresponds to an
uncertain measurement (see text).}
\label{fig:corr_pmax_bi}
\end{figure}

\begin{figure}
\resizebox{\hsize}{!}{\rotatebox{0}{\includegraphics{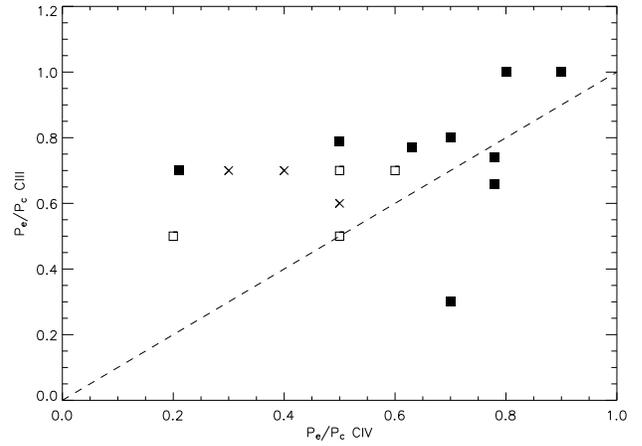}}} 
\caption[]{The correlation between the polarization degrees in the \civ\, and
\ciii\, emission lines relative to the polarization in the adjacent redward
continuum.  The dashed line refers to an equal polarization in the two
lines.  Symbols are identical to those used in 
Fig. \ref{fig:corr_pmax_bi}.}
\label{fig:corr_pe1_pe2}
\end{figure}

Although the relation $p_{\rm max}$ $-$ BI (see Fig. \ref{fig:corr_pmax_bi}) is the most 
significant correlation in Table \ref{table:spec_res}, we must consider it with caution.
Indeed, $p_{\rm max}$ being measured where the absorption is the deepest, it may depend on
the spectral resolution, more particularly when the residual intensity in the trough is nearly
zero, which usually corresponds to large BI.  In this case, it is possible that the higher the
resolution, the lower the flux in a single pixel and the larger the measured $p_{\rm max}$.
However, a majority of BAL QSOs with large BI do not have completely black troughs, and the
$p_{\rm max} - {\rm BI}$ correlation is most probably real.  This correlation is discussed 
in more details in Sect. 6.2.

As found in previous spectropolarimetric studies (e.g. Ogle et al. \cite{OGL99}), the polarization 
in the emission lines is systematically lower than that the polarization in the continuum, which
corresponds to $p_e/p_c \leq 1$ for both the \civ\, and \ciii\, emission lines (see Table 
\ref{table:specmeas}). In addition, we find in Table \ref{table:spec_res} another significant 
correlation which relates the polarization in the \civ \, and \ciii \, emission lines. As 
illustrated in Fig. \ref{fig:corr_pe1_pe2}, the polarization in the \ciii \, emission 
line is not only correlated to but also systematically higher than the polarization in the \civ \, 
emission line.  If the broad emission lines appear in the polarized flux, one possible explanation 
may be the effect of two different polarizing mechanisms, one of them either depolarizing \civ\, 
or polarizing \ciii. Lee (\cite{LEE94}) has shown that resonance scattering is more efficient at 
producing polarization in semi-forbidden lines such as \ciii \, than in most permitted lines 
such as \civ.  On the other hand, if there is only one polarizing mechanism at work for both 
emission lines, the emitting regions may have different geometries and/or sizes with respect 
to the scattering region (e.g. Goodrich \& Miller \cite{GOO95}).  Cassinelli et al. (\cite{CAS87}) 
have calculated the polarization of a spherical emission region of radius R scattered by a region 
located at a distance $D > R$.  They find that the polarization should be corrected by a factor 
$\sqrt{1-(R/D)^2}$ compared to the case of a point source.  If the size of the \ciii\, 
emitting region is smaller than the size of the \civ\, emitting region, the geometric dilution
may induce a larger polarization in the \ciii\, emission line than in the \civ\, emission line.
Of course, one must keep in mind that neither the broad emission line region nor the scattering 
region are likely to be spherical. Since these results are in contradiction with reverberation 
mapping studies (Clavel et al. \cite{CLA91}, Chiang \& Murray \cite{CHI96}), this explanation 
based on a purely geometrical dilution seems unlikely.  In some BAL QSOs, the broad emission lines 
do not appear in the polarized flux indicating that they are not polarized at all.  In this case, 
the observed correlation may simply reflect the fact that the polarization in the continuum is 
larger in the blue part of the spectrum (i.e. at the wavelength of \civ) than in the red part
(i.e. at the wavelength of \ciii).

\begin{figure}
\resizebox{\hsize}{!}{\rotatebox{0}{\includegraphics{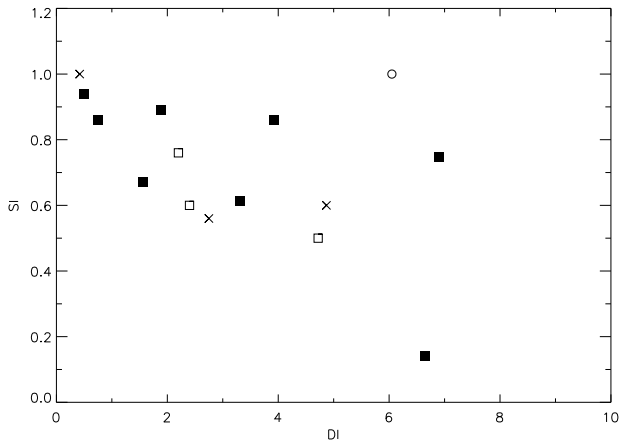}}} 
\caption[]{The correlation between the detachment index DI and SI, the ratio of 
the \civ \, absorption depths in the polarized flux and in the total 
flux.  Both quantities are unitless.  Symbols are identical to those used in 
Fig. \ref{fig:corr_pmax_bi}.  The open circle corresponds to an uncertain 
measurement  (see text)}
\label{fig:corr_di_si}
\end{figure}

There is also a marginal ($P \simeq 0.03$) correlation between the detachment index and
SI, which is worth considering.  In Fig. \ref{fig:corr_di_si}, we plot SI
versus DI.  It clearly appears that if we do not consider the object represented by an
open circle\footnote{By carefully inspecting the data of Ogle et al. (\cite{OGL99}), we noticed 
that the \civ \, absorption line in \object{B1212$+$1445} is partly cut at the lower end of the 
spectrum such that the value of SI given in Table \ref{table:spec_res} for this object 
might be highly uncertain.}, the correlation between SI and DI
is much more significant.  Indeed, in this case, the Kendall $\tau$
and the Spearman $r_s$ statistics give a probability lower than 0.005
that these two quantities are correlated by chance.  Additional
measurements are badly needed to confirm this possible correlation.
The correlation between SI and DI indicates that
in P Cygni type BAL QSOs, the BALs in the polarized flux are nearly as
deep as in the direct flux (cf. \object{B2225$-$0543}), while in BAL QSOs with
detached troughs, the BALs in the polarized flux are much shallower than
in the direct flux (cf. \object{B1246$-$0542}).  As discussed in the next
section, the consequences of such a relation for constraining BAL QSO
models may be important.

\section{Discussion}

\subsection{The principal component PC2 : the anti-correlation $p_0$ - DI}

The fact that $p_0$ is only correlated with the detachment index emphasizes 
the importance of DI to understand the geometry of the outflow in BAL QSOs.  
The PCA has confirmed the importance of the $p_0-{\rm DI}$ anti-correlation 
which is dominant in PC2, one of the two significant principal components.

The WfD model of Murray et al. (\cite{MUR95}) proposes a natural explanation of DI, hence 
of the detached troughs observed in some BAL QSOs (see Fig. \ref{fig:wfd}) .  According to 
this model, an observer looking through the wind close to the disk (edge-on) sees 
absorption at all velocities including at low velocity. P~Cygni type profiles are 
then produced and the observer measures a small value of DI.  On the contrary, if the 
line-of-sight towards the observer grazes the upper edge of the wind, 
it misses the low velocity part of the absorption, the wind streamlines being more 
vertical than radial at low velocities.  Therefore the absorption may start 
several thousand kilometers per second blueward of the corresponding emission peak, 
resulting in detached troughs and large values of DI.  In the WfD model, DI is 
naturally related to the orientation of the accretion disk with respect to the line
of sight.  On the other hand, in the torus model (Schmidt \& Hines \cite{SCH97}),
DI has no simple explanation. In polar wind models such as the one described by 
Punsly (\cite{PUN99}), the detached troughs are caused by intermittent variations of matter 
ejections along the bipolar outflow.  In this case, the line detachment is related to a 
time-dependent phenomenon and does not depend on orientation.

Since the continuum polarization may also depend on the inclination, the relation between 
$p_0$ and DI favours a BALR model which is orientation dependent.  In the WfD model, the 
dependence of $p_0$ with DI is naturally explained since higher polarization is expected 
close to the disk where attenuation of the direct unpolarized continuum is larger.  As the 
inclination (measured from the disk axis) decreases, the growing importance of the direct 
continuum gives rise to a dilution of the polarization.  Therefore, detached troughs are 
naturally associated with lower optical polarization. Since the polarization is not correlated 
with the slope of the continuum, it rules out an extinction by small dust particles as being
the dominant extinction mechanism.  A significant part of the attenuation should then be neutral, 
dominated either by electron scattering or by extinction by large dust particles.  

The scatterers at the origin of the polarization may lie in the inner parts of the wind itself 
or they may be located in a polar region, which is usually assumed to interpret the 
spectropolarimetric observations (e.g. Hines \& Wills \cite{HIN95}, Goodrich \& Miller 
\cite{GOO95}), although its origin is unclear.

If the scatterers are located in the wind, the electron scattering model within the cylindrical 
sector geometry depicted by Brown \& Mc Lean (\cite{BRO77}) may be considered to explain the 
$p_0-{\rm DI}$ anti-correlation (see also Paper I).  More particularly, as discussed in Sect. 3, 
Fig. \ref{fig:corr}a shows that the relation $p_0-{\rm DI}$ is not a true anti-correlation : for 
low values of DI (P Cygni profiles), there is a large range of $p_0$ values.  At a given 
inclination, the Brown \& Mc Lean (\cite{BRO77}) model predicts 
a polarization which increases with the half-opening angle of the wind.  The range in $p_0$ for 
a given inclination may therefore simply reflect a range of opening angles of the equatorial wind. 
For large values of DI (profiles with detached troughs), the strong dilution by the direct 
unpolarized continuum leads to small values of $p_0$ whatever the opening angle of the wind. The 
larger polarization of LIBAL QSOs compared to HIBAL QSOs (Paper I, Lamy \cite{LAM03}) may then be 
explained either by the stronger extinctions observed in LIBAL QSOs (Sprayberry \& Foltz 
\cite{SPR92}, Reichard et al. \cite{REI03}) or by a larger opening angle of their wind.  This 
latter possibility is in agreement with the high covering factors needed to interpret their very 
weak \oiii\, emission (Boroson \& Meyers \cite{BOR92b}, Yuan \& Wills \cite{YUA03}).  This simple 
explanation qualitatively reproduces the observed $p_0 - {\rm DI}$ correlation. For HIBAL QSOs, the 
correlation is possibly weaker because of the variation from object to object such that the smaller 
range of $p_0$ values may mask the correlation.

\subsection{The anti-correlation SI $-$ DI and the correlation $p_{\rm max} - {\rm BI}$}

\begin{figure}
\resizebox{\hsize}{!}{\rotatebox{0}{\includegraphics{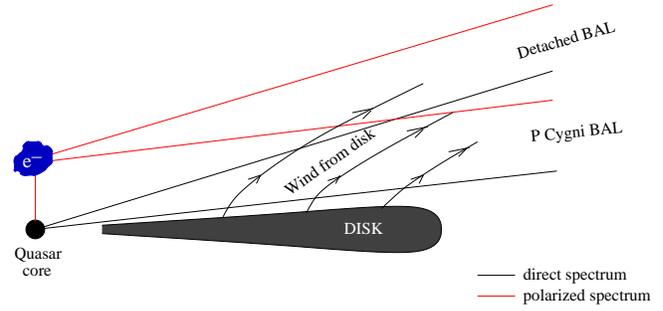}}} 
\caption[]{Schematic illustration of the SI-DI relationship.  The equatorial wind, represented
by the curved streamlines, emerges from a geometrically thin accretion disk located in the 
equatorial plane.  A mirror of electrons is located in the polar region and scatters some
radiation coming from the central regions along the line-of-sight.  This polarized component
is then subsequently absorbed in the equatorial wind.  When the disk is seen edge-on, P Cygni 
type profiles are observed since the observer sees absorption at all velocities including the low 
ones near the disk.  For less inclined line-of-sights, detached absorption profiles are observed 
since the observer misses the low velocity part of the absorption.}
\label{fig:wfd}
\end{figure}

If the scatterers are located in a polar region, part of the light from the continuum may be 
scattered and subsequently absorbed in the equatorial wind.  The variation of SI with DI, derived 
from the spectropolarimetric data, is naturally explained within this geometry, as illustrated in
Fig. \ref{fig:wfd} : for nearly equatorial line-of-sights (DI $\ll$), the scattered flux crosses
regions of the equatorial wind located close to the accretion disk where the opacity in the
absorption lines is large.  Therefore the polarized flux is significantly absorbed and shows
deep broad absorption lines (SI $\sim 1$, cf. \object{B2225$-$0534}). For less inclined 
line-of-sights (DI $\gg$), the scattered flux crosses regions of the equatorial wind where the 
opacity is much lower resulting in weaker absorptions in the polarized flux (SI $\ll$).  The 
quantity SI becomes $\simeq 0$ in the extreme case where the scattered flux is not absorbed at 
all (cf. \object{B1246$-$0542}).  
It is more difficult to understand this relation if the scatterers are located in the equatorial 
wind itself than in a polar region.  The SI-DI relationship appears therefore important for 
constraining the geometry of the BALR, and is definitely worth confirming with additional 
spectropolarimetric observations.

The other correlation derived from the spectropolarimetric data, $p_{\rm max} - {\rm BI}$,
may also be simply explained with such a model.  If BI is large, the direct unpolarized flux is 
strongly absorbed in the line resulting in a smaller dilution of the scattered component at these 
wavelengths. Therefore large values of $p_{\rm max}$ are expected in the case of strong 
absorptions.

\subsection{A ``two-component wind'' model ?}

Although the $p_0 - {\rm DI}$ and SI $-$ DI anti-correlations can be interpreted assuming an
equatorial wind issued from a disk, the existence of an additional polar region where scattering
takes place is apparently needed, at least in some BAL QSOs.  This additional region may simply 
consist of another part of the outflow with a lower density and a larger amount of free electrons.  
Since the properties of these possibly related regions are different, we call hereafter this model 
the ``two-component wind'' (2CW) model, emphasizing that it is a natural extension or a variant of 
the WfD model.

The 2CW model is supported by several theoretical models which have been developed in the last years 
(Proga et al. \cite{PRO98}, \cite{PRO99}, Proga \cite{PRO00}, \cite{PRO03}, Pereyra et al. 
\cite{PER04}).  These authors generalize the Murray et al. model, in particular by including a 
magnetic field distribution.  The outflow may then be radiation or magnetically driven or both.  
The simulations show that different types of winds can be produced 
(from being nearly polar to nearly equatorial) by varying the magnetic field,
the total luminosity or the ratio between the luminosity of the disk and that of the central 
compact object.  In particular, one may qualitatively reproduce a 2CW model with a dense
equatorial wind bounded on the polar side by a much less dense stream. 

Observationally, the rotation of the polarization angle $\theta$ accross broad absorption 
lines (a remarkable instance is given in \object{B1413$+$1143}, see Fig. \ref{fig:spectropola2}) may
provide evidence for the existence of two scattering regions in some BAL QSOs.  Indeed, 
the continuum may be polarized by scattering from a polar and an equatorial distribution of
electrons (or dust) at the same time. The scattered flux from the polar region has a polarization 
angle perpendicular to the disk axis while the polarization produced in the disk is parallel.  
The resulting polarization angle depends on the relative importance of both components.  
If the scattered flux from the polar region is re-absorbed in the equatorial wind, its contribution 
is reduced at these wavelengths leading to the observed rotation of the polarization angle.   On 
the other hand, there may be only one source of scatterers, and the rotation of $\theta$ may result
from another polarizing mechanism, for example resonance scattering since in P Cygni type profiles
the broad absorption is partly filled with resonantly scattered emission. 

Besides polarization, another indicator of geometry is the radio morphology.  Recently, 
some observational evidence supporting the existence of much smaller and weaker
radio jets in radio-quiet QSOs have recently emerged (Kuncic \cite{KUN99}).  Since the axis of 
the accretion disk is usually thought to be aligned with the radio jets, the comparison
of the position angles of radio axes ($PA$) and optical polarizations ($\theta$)
can in principle give clues on the location of the scattering region. If the broad 
absorption lines occur in an equatorial wind, BAL QSOs are quasars seen edge-on and we 
should detect a radio axis close to the plane of the sky.  The optical polarization should be 
roughly parallel to the radio axis if the scatterers are located in the wind or roughly 
perpendicular to it if they lie in a polar region.  Unfortunately, because of their weak 
radio fluxes and their large distance, there are very few BAL QSOs with a resolved radio 
structure. In the low redshift BAL QSO \object{PG 1700$+$5153}, Kellermann et al. (\cite{KEL94}) 
measure $PA \sim 145^{\circ}$ for the radio structure observed at the VLA while Ogle et al. 
(\cite{OGL99}) and Schmidt \& Hines (\cite{SCH99}) measure a polarization position angle $\theta 
\sim 53^{\circ}$.  In the radio-loud BAL QSO \object{LBQS 1138$-$0126}, Brotherton et al. 
(\cite{BRO02}) observe a double-lobed radio structure 
with $PA \sim 52^{\circ}$ characteristic of an edge-on source in unified radio schemes.  Their 
spectropolarimetry of \object{LBQS 1138$-$0126} shows a continuous rotation of the continuum position 
angle from $180^{\circ}$ in the blue to $150^{\circ}$ in the red part of their optical spectra. 
In both objects, the optical polarization is then more or less perpendicular to the radio structure,
suggesting that the scatterers are located in a polar region.  However, we must keep in mind 
that at the VLA resolution, the weak radio axis may be twisted by some interactions with the 
surrounding medium.  Therefore, higher resolution images are needed to confirm the VLA results.  
Recently, Jiang \& Wang (\cite{JIA03}) performed the first VLBI observations of three BAL QSOs.  
Among them, \object{J\,1312$+$2319} has a two-sided structure with a position angle $PA \sim 
59^{\circ}$ indicating an edge-on orientation.  We have recently obtained the optical polarization 
of this object for which we find $\theta \sim 166^{\circ}$ (Sluse et al. 2004) indicating that the 
two position angles are roughly perpendicular.   Although there are only very few radio observations 
of BAL QSOs, they apparently support a 2CW model, where scattering occurs in a polar region.

In the 2CW model, the BALR is located in the equatorial plane.  However, based on radio 
observations of FIRST BAL QSOs with the VLA, Becker et al. (\cite{BEC00}) argue that BAL QSOs are 
not viewed from any particular orientation because they observed both steep and flat spectra within 
a sample of BAL QSOs discovered in the FIRST catalog.  While, in BAL QSOs, it is not sure 
that the radio spectrum is a good indicator of the position angle of the radio axis as it is in 
radio-loud sources, the recent VLBI observations of Jiang \& Wang (\cite{JIA03}) may confirm the
claim of Becker et al. (\cite{BEC00}).  Indeed, while \object{J\,1312$+$2319} is seen edge-on, 
\object{J\,1556$+$3517} has a compact structure and a flat radio spectrum suggesting it is 
probably seen pole-on.  However, \object{J\,1556$+$3517} has also very unusual spectropolarimetric 
properties (Brotherton et al. \cite{BRO97}), possibly making it a peculiar object.

\subsection{The principal component PC1 : the correlation BI - \feii}

Until now, we have concentrated on PC2 which is most probably driven by orientation.  But what 
drives PC1 in BAL QSOs~?  Yuan \& Wills (\cite{YUA03}) have shown that $z \sim 2$ BAL QSOs follow the 
main correlations linking the properties of emission lines as depicted in the so-called Boroson 
\& Green eigenvector (\cite{BOR92}).  In particular, BAL QSOs have the strongest \feii \, and weakest 
\oiii \, emission among high redshift quasars.  For non-BAL QSOs, these properties are usually 
interpreted in terms of accretion rates of gas around the central massive black hole 
(Boroson \& Green \cite{BOR92}). When the accretion rate increases, the disk inflates because of the 
larger radiation pressure. The optical depth becomes higher and the optically thick gas 
prevents the UV ionizing photons to reach the narrow-line region where \oiii \, forms.  At 
the same time, the \feii \, emission strongly increases because of a larger amount of X-rays.  
In this scenario, BAL QSOs are quasars which accrete matter with a rate close to the Eddington 
rate (Yuan \& Wills \cite{YUA03}) such that they display the most extreme \feii \, and \oiii\, emission 
properties.  In the WfD model, it is possible that the larger the accretion rate, the higher the gas density in the 
wind and therefore the stronger the absorption. This could possibly explain the positive 
correlation observed between BI and \feii \,$\lambda$ 2400, which dominates PC1.

\section{Conclusions}

With new broad-band polarization data and new measurements of indices on good quality 
spectra from the literature, we have performed a systematic search for correlations 
among a sample of 139 BAL QSOs. We find six non-trivial correlations.  We use a principal 
component analysis to isolate the main correlations in only two principal components : 
PC1, dominated by the BI $-$ \feii\, correlation, is possibly related to the accretion 
rate of matter around the central massive black hole, while PC2, dominated by the 
$p_0 -{\rm DI}$ anti-correlation, could be driven by orientation.  

We have also searched for correlations among indices measured on spectropolarimetric data for a 
sample of 21 BAL QSOs.  We find that the polarization in the \ciii\,emission line is 
systematically larger than the polarization in the \civ\,emission line, and that the maximum of 
the polarization in the \civ\, absorption line is higher for BAL QSOs with large values of BI.  
Another important result is a possible correlation between the detachment index 
DI and SI, a quantity which measures the depth of the \civ\, BAL in the polarized flux relative 
to that one in the total flux. These statistical studies outline the importance of polarization 
to disentangle the geometry of the outflows in BAL QSOs. We have also emphasized that BI
and more particularly DI are critical parameters to describe the BAL phenomenon.  

The anti-correlations $p_0-{\rm DI}$ and DI$-$SI are naturally explained with a model where 
the continuum attenuation and the line absorption occur within a continuous equatorial wind 
accelerated from a disk. The direct continuum is also assumed to be scattered by a ``mirror'' of 
electrons located in polar regions, possibly belonging to a polar wind. For edge-on line-of-sights, 
the direct unpolarized continuum is attenuated in the dense equatorial regions near the disk. 
Therefore, in BAL QSOs with P Cygni profiles, a large continuum polarization is observed because 
the relative contribution of the scattered polarized component is large.  The scattered flux is 
also absorbed in the equatorial wind and a prominent absorption line is observed in the
polarized spectrum.  For a less inclined line-of-sight grazing the upper edge of the 
equatorial wind, the direct unpolarized continuum is less heavily absorbed and dilutes the 
continuum polarization.  The scattered flux is also less absorbed in these lower opacity 
regions of the wind such that for BAL QSOs with detached absorptions, lower continuum polarization 
and shallow troughs in the polarized flux are observed.

However, with polarization data only, we cannot unambiguously locate the scatterers. 
A convenient way to solve this problem is to compare the polarization position angle
with the orientation of the radio-axis.  Unfortunately, only 3 BAL QSOs have been
resolved at the radio wavelengths and have polarimetric data available.  Interestingly, in 
these 3 objects, the optical polarization angle is nearly perpendicular to the 
radio-axis, as expected from a model where scattering occurs in a polar region.  
 
\begin{acknowledgements}
H. Lamy thanks Prof. J. Lemaire and the Belgian Institute of Space Aeronomy for giving him 
the opportunity of performing observations in La Silla (Chile) in March 2002.  The authors
would also like to thank the referee for the careful reading.
\end{acknowledgements}


\end{document}